\documentclass[a4paper,12pt]{article}
\usepackage[centering,hmargin=2.0cm,vmargin=2.5cm]{geometry}
\usepackage{amsmath, amsfonts, amssymb}
\usepackage{graphicx}
\usepackage{enumerate}
\usepackage{hyperref}
\usepackage{float}
\usepackage[font={small},textfont={it}]{caption} 
\usepackage{cite}
\usepackage[table,xcdraw,dvipsnames]{xcolor}
\usepackage{multirow,multicol}
\hypersetup{colorlinks,bookmarksopen,bookmarksnumbered,linkcolor=Scarlet,pdfstartview=FitH,urlcolor=Scarlet,citecolor=Scarlet}
\allowdisplaybreaks
\usepackage{color}
\usepackage[toc,page]{appendix}
\usepackage[bottom]{footmisc}
\definecolor{Scarlet}{cmyk}{0,1,1,0.55}

\newcommand\mpl{M_{\rm Pl}}
\newcommand{\g}{g_{a\gamma \gamma}}
\newcommand{\FF}{F_{\mu \nu} F^{\mu \nu}}
\newcommand{\dFF}{F_{\mu \nu} {}^{*}F^{\mu \nu}}

\begin{document}
\begin{center}  
{\huge\bf\color{Scarlet} Spinning Black Holes with Axion Hair} \\[3ex]
{\bf\large Clare Burrage$^{a}$\footnote{clare.burrage@nottingham.ac.uk}, Pedro G. S. Fernandes$^{a}$\footnote{pedro.fernandes@nottingham.ac.uk}}\\[2ex]
{\it $^a$ School of Physics and Astronomy, University of Nottingham, University Park, Nottingham, NG7 2RD, United Kingdom}\\[3ex]
{\bf\large Richard Brito$^{b}$\footnote{richard.brito@tecnico.ulisboa.pt}, Vitor Cardoso$^{bc}$\footnote{vitor.cardoso@nbi.ku.dk}}\\[2ex]
{\it $^b$ CENTRA, Departamento de Física, Instituto Superior Técnico – IST, Universidade de Lisboa – UL, Avenida Rovisco Pais 1, 1049–001 Lisboa, Portugal}\\[3ex]
{\it $^c$ Niels Bohr International Academy, Niels Bohr Institute, Blegdamsvej 17, 2100 Copenhagen, Denmark}\\[3ex]

{\large\bf Abstract}\begin{quote}
In this work we construct and analyse non-perturbative stationary and axially-symmetric black hole solutions in General Relativity coupled to an electromagnetic and an axion field. The axion field is coupled to the electromagnetic field, which leads to hairy solutions in the presence of an electric charge and rotation. We investigate the existence and characteristics of these solutions for different values of the spin, charge and coupling constant. Our analysis shows the presence of violations of the Kerr-Newman bound, solutions with large positive and negative values of the gyromagnetic ratio, and the existence of multiple branches of solutions with distinct properties, demonstrating that black hole uniqueness does not hold in this scenario. The code used in this study is \href{https://github.com/pgsfernandes/SpinningBlackHoles.jl}{publicly available}, providing a valuable tool for further research on this model.
\end{quote}
\end{center}

\section{Introduction}
The existence of dark matter, which is thought to make up around 85\% of the universe's mass, is one of the greatest mysteries in modern astrophysics. A promising dark-matter candidate are axions, first proposed by Peccei and Quinn as a potential solution to the strong CP problem \cite{Peccei:1977hh}. They can be extremely light and weakly interacting \cite{Bergstrom:2009ib}. Additionally, axion-like fields have been observed to arise naturally in string theory constructions \cite{Svrcek:2006yi, Arvanitaki:2009fg}, further adding to the motivation for studying their potential consequences.

In the strong gravity regime, massive scalar fields, such as axions, are expected to trigger superradiant instabilities around spinning black holes \cite{Detweiler:1980uk,Dolan:2007mj,Cardoso:2005vk,East:2017ovw,East:2018glu,Brito:2015oca,Witek:2012tr}. This causes the black hole to lose some of its rotational energy, which is then transferred to a dense cloud of axions. Over time, the cloud will emit gravitational waves and slowly lose energy, causing it to shrink and eventually disappear. These gravitational waves, as well as a plethora of other unique signatures, can be detected by current and future instruments \cite{Arvanitaki:2010sy,Arvanitaki:2014wva,Brito:2014wla,Arvanitaki:2016qwi,Baryakhtar:2017ngi,Brito:2017wnc,Brito:2017zvb,Hannuksela:2018izj,Baumann:2018vus,Baumann:2021fkf}, making these systems an interesting subject for study. Moreover, it has been proposed that when photons interact with superradiant axion clouds, they can produce sudden, intense bursts of light that are similar to the emissions from a laser at a quantum level \cite{Rosa:2017ury,Sen:2018cjt,Ikeda:2018nhb,Boskovic:2018lkj}. 

No-hair theorems state that electrovacuum, stationary black holes in Einstein-Maxwell theory are described by the Kerr-Newman solution~\cite{Mazur:1982db} (see also Refs.~\cite{Mazur:2000pn,Robinson:2004zz} for reviews). However, if couplings are allowed between the scalar and electromagnetic fields, black holes can have additional properties, known as ``hair". These hairy black hole solutions have long been studied in contexts such as Kaluza-Klein theory and supergravity \cite{Gibbons:1987ps,Garfinkle:1990qj}. More recently, couplings of this type and respective black hole solutions have been explored in the context of black hole spontaneous scalarization \cite{Herdeiro:2018wub,Fernandes:2019rez,Fernandes:2020gay,Fernandes:2019kmh,Boskovic:2018lkj,Lin:2023npr} (see Ref. \cite{Doneva:2022ewd} for a recent review).

Previous studies have investigated the effects of axions on magnetically charged static and spherically symmetric black holes\footnote{In this scenario, the system's symmetries result in the field equations being equivalent to a set of ordinary differential equations, rather than partial differential equations.} \cite{Fernandes:2019kmh,Lee:1991jw,Filippini:2019cqk} and perturbative solutions away from the slowly-rotating Kerr-Newman black hole \cite{Boskovic:2018lkj,KIczek:2021vlc}. In this work, we go beyond the perturbative/magnetically charged framework and construct fully non-linear stationary and axially-symmetric solutions of electrically charged and rotating black hole spacetimes with axionic hair. Unlike uncharged and/or spherically symmetric cases without magnetic charge, the axion is sourced by a non-zero electromagnetic field in this scenario, resulting in equilibrium hairy configurations.

The structure of this paper is as follows. In Section \ref{sec:setup}, we present the model, along with the equations of motion and the ansatz used for solving them. We also explain the boundary conditions applied and our numerical methodology. Section \ref{sec:results} contains the main outcomes of our study. Subsection \ref{sec:smallspin} deals with the solutions obtained for small spin values, while subsection \ref{sec:highspin} presents the results for higher spin values. We summarize our findings in Section \ref{sec:conclusions}.
The source code utilized to construct the solutions examined in this work is accessible to the public \cite{gitlink}, enabling further investigations into this model.

\section{The model and setup}
\label{sec:setup}
\begin{figure}[]
    \centering
    \includegraphics[width=0.75\textwidth]{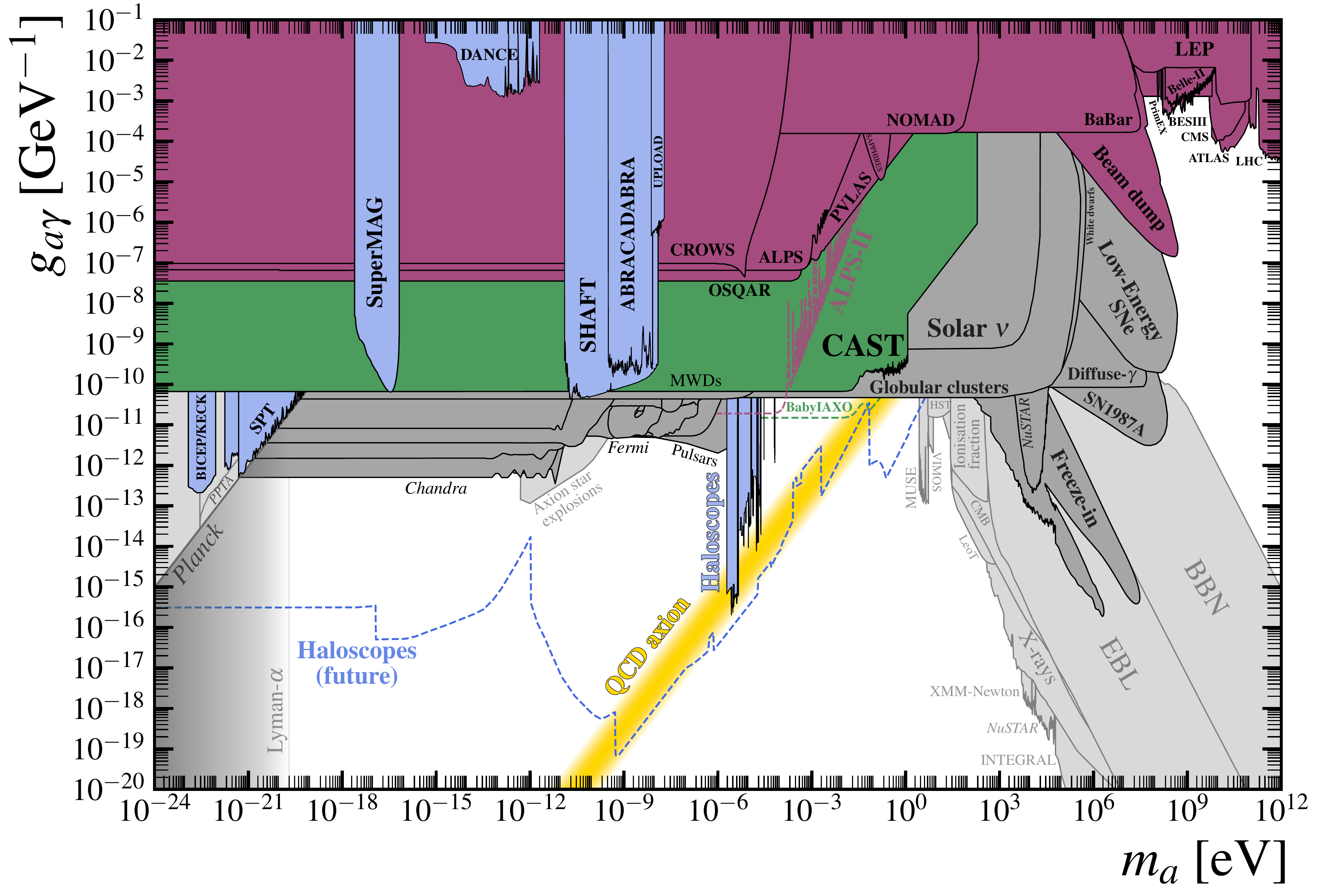}
    \caption{The figure shows the constraints on the axion coupling constant $\g$ and mass $m_a$ from various experiments. The QCD axion is represented by the yellow band. The magenta band represents constraints from collider physics experiments, the blue from dark matter experiments, the green from stellar experiments, and the gray from astrophysics. The dashed lines indicate projections from future observations. Adapted from Ref. \cite{AxionLimits}.}
    \label{fig:axion_constraints}
\end{figure}

We will explore Einstein's gravity coupled to one of the best-motivated extensions of the Standard Model, which is described by the following action
\begin{equation}
    S = \int d^4x \sqrt{-g} \left[\frac{\mpl^2}{2} R - \frac{1}{4} \FF - \frac{1}{2} \partial_\mu a \partial^\mu a - \frac{1}{2}m_a^2 a^2 - \frac{\g}{4} a \dFF \right],
    \label{eq:action}
\end{equation}
where ${}^{*}F^{\mu \nu} = \frac{1}{2\sqrt{-g}} \epsilon^{\mu \nu \alpha \beta} F_{\alpha \beta}$ is the dual of the U(1) gauge field strength, $F_{\mu \nu} = \partial_\mu \mathcal{A}_\nu - \partial_\nu \mathcal{A}_\mu$, $\epsilon^{\mu \nu \alpha \beta}$ the totally anti-symmetric Levi-Civita symbol, and the axion $a$ is a real pseudoscalar field with mass $m_a$ that couples to electromagnetism. The axionic coupling constant $\g$, has dimensions of inverse mass. The currently allowed parameter space for the axion model described by action \eqref{eq:action} is shown in Fig. \ref{fig:axion_constraints}, in the $(m_a,\g)$ plane. In the absence of a mass term, the axion enjoys the shift-symmetry $a\to a+c$, for constant $c$.

Varying the action \eqref{eq:action} with respect to the metric, we obtain the field equations
\begin{equation}
    R_{\mu \nu}-\frac{1}{2} g_{\mu \nu}R = \frac{1}{\mpl^2} T_{\mu \nu},
    \label{eq:EinEqs}
\end{equation}
where the total stress-energy tensor is given by
\begin{equation}
    T_{\mu \nu} = \partial_\mu a \partial_\nu a - \frac{1}{2}g_{\mu \nu} \left[ \partial_\alpha a \partial^\alpha a + m_a^2 a^2 \right] + F_{\mu}^{\phantom{\mu} \alpha} F_{\nu \alpha} - \frac{1}{4} g_{\mu \nu} F_{\alpha \beta} F^{\alpha \beta},
\end{equation}
which is unaffected by the axionic coupling. The axion equation of motion is
\begin{equation}
    \Box a = m_a^2 a + \frac{\g}{4} \dFF,
    \label{eq:axionEq}
\end{equation}
and the Maxwell-axion equations are
\begin{equation}
    \nabla_\nu \left(F^{\mu \nu} + \g a\, {}^{*}F^{\mu \nu} \right) = 0.
    \label{eq:MaxwellEqs}
\end{equation}
From now on, for convenience, we work in units of $\sqrt{2}\mpl$ by performing the rescalings
\begin{equation}
    a \to \sqrt{2}\mpl a, \quad \mathcal{A}_\mu \to \sqrt{2}\mpl \mathcal{A}_\mu, \quad \g \to \frac{\g}{\sqrt{2}\mpl},
    \label{eq:units}
\end{equation}
which amount to replacing $1/\mpl^2 \to 2$ in Eq. \eqref{eq:EinEqs}, while Eqs. \eqref{eq:axionEq} and \eqref{eq:MaxwellEqs} remain unchanged. Furthermore, in this work we consider only a massless axion for simplicity. However, we have performed initial simulations for a massive axion and found that the results are qualitatively similar.

We aim to study black hole solutions to the equations of motion \eqref{eq:EinEqs}, \eqref{eq:axionEq}, \eqref{eq:MaxwellEqs}, that are regular on and outside of the event horizon, stationary and axially-symmetric. These solutions possess two commuting Killing vector fields, $\xi = \partial_t$ and $\eta = \partial_\varphi$, in an adapted coordinate system. As such, we consider the following metric ansatz in quasi-isotropic coordinates:
\begin{equation}
\begin{aligned}
    &ds^2 = -f \mathcal{N}^2 dt^2 + \frac{g}{f} \left[ h \left(dr^2 + r^2 d\theta^2\right) + r^2 \sin^2\theta \left(d\varphi - \frac{W}{r}\left(1-\mathcal{N}\right) dt\right)^2\right],\\&
    \mathrm{where} \quad \mathcal{N} \equiv \mathcal{N}(r) = 1-\frac{r_H}{r},
\end{aligned}
\label{eq:metric}
\end{equation}
and where $f$, $g$, $h$, $W$ are dimensionless functions of $r$ and $\theta$, and $r_H$ is the coordinate location of the event horizon. This ansatz for the metric is motivated by discussions such as those in Ref. \cite{Xie:2021bur}, and the Kerr-Newman solution can be written in this form (see Appendix \ref{ap:KN}). For the electromagnetic four-potential, we consider the following ansatz:
\begin{equation}
    \mathcal{A}_\mu dx^\mu = \left(A_t - \frac{W}{r}\left(1-\mathcal{N}\right) A_\varphi \sin^2 \theta \right) dt + A_\varphi \sin^2 \theta d\varphi,
    \label{eq:ansatzEM}
\end{equation}
where $A_t$ and $A_\varphi$ depend on both $r$ and $\theta$, as does the axion $a$.

To solve the system, we follow Ref. \cite{Fernandes:2022gde} and use the following combination of field equations \eqref{eq:EinEqs} (written here in the form $\mathcal{E} = 0$):
\begin{equation}
  \begin{aligned}
    &-\mathcal{E}^{\mu}_{\phantom{\mu} \mu} + 2 \mathcal{E}^{t}_{\phantom{t} t} + \frac{2W r_H}{r^2} \mathcal{E}^{\varphi}_{\phantom{\varphi} t} = 0,\\&
    \mathcal{E}^{\varphi}_{\phantom{\varphi} t} = 0,\\&
    \mathcal{E}^{r}_{\phantom{r} r} + \mathcal{E}^{\theta}_{\phantom{\theta} \theta} = 0,\\&
    \mathcal{E}^{\varphi}_{\phantom{\varphi} \varphi} - \frac{W r_H}{r^2} \mathcal{E}^{\varphi}_{\phantom{\varphi} t} - \mathcal{E}^{r}_{\phantom{r} r} - \mathcal{E}^{\theta}_{\phantom{\theta} \theta} = 0,
  \end{aligned}
  \label{eq:feqs_combination}
\end{equation}
together with the axion equation \eqref{eq:axionEq} and the two non-trivial components of the Maxwell equations \eqref{eq:MaxwellEqs}.
It is worth noting that the third equation in Eq. \eqref{eq:feqs_combination} is equivalent to an equation that, for a massless axion ($m_a=0$), is exactly the same as in vacuum General Relativity and depends only on the metric function $g$. Using an ansatz of the form $g=g_r(r) g_\theta(\theta)$, the equation is separable and, by imposing regularity, we find that the angular part is trivial, leaving $g = \left(1+\frac{r_H}{r}\right)^2$, as is the case for a Kerr-Newman black hole. Although we know a closed-form solution for the metric function $g$, we will not use it in the numerical method and will instead use it as a test for the code.

To solve the problem numerically we compactify our radial coordinate as
\begin{equation}
    x = 1-\frac{2r_H}{r},
\end{equation}
mapping $r\in [r_H, \infty[$ to $x \in [-1,1]$. We impose the following boundary conditions. Axial symmetry and regularity on the axis of symmetry $\theta=0, \pi$, impose
\begin{equation}
  \partial_\theta f = \partial_\theta g = \partial_\theta h = \partial_\theta W = \partial_\theta a = \partial_\theta A_t = \partial_\theta A_\varphi = 0, \qquad \mathrm{for} \quad \theta=0, \pi.
\end{equation}
The involved functions possess definite parity with respect to $\theta=\pi/2$ (all are even parity, except $a$ which has odd parity), and therefore we need only consider the range $\theta \in [0,\pi/2]$. Parity considerations then imply
\begin{equation}
      \partial_\theta f = \partial_\theta g = \partial_\theta h = \partial_\theta W = \partial_\theta A_t = \partial_\theta A_\varphi = 0, \qquad a = 0, \qquad \mathrm{for} \quad \theta=\pi/2.
\end{equation}
Asymptotic flatness is imposed by the boundary conditions
\begin{equation}
\begin{aligned}
    &f=g=h=1,\qquad
    \partial_x W + \chi \left(1 + \partial_x f\right)^2 = 0,\\&
    a = 0,\\&
    \partial_x A_t -\frac{q}{2} \left(1 + \partial_x f\right) = 0,\qquad A_\varphi = 0,
\end{aligned}
\end{equation}
for $x=1$, where
\begin{equation}
    \chi \equiv \frac{J}{M^2}, \qquad q \equiv \frac{Q}{M},
    \label{eq:dim_quantities}
\end{equation}
are the dimensionless spin and charge, respectively, with $M$, $J$, and $Q$ being the ADM mass, angular momentum, and electric charge respectively. These can be extracted from the asymptotic behaviour discussed below in Eq. \eqref{eq:asymptotic}. At the horizon we impose
\begin{equation}
    \begin{aligned}
        &f-2 \partial_x f = 0, \qquad
        g+2 \partial_x g = 0,\qquad 
        \partial_x h = 0,\qquad 
        W - \partial_x W = 0,\\&
        \partial_x a = 0,\\&
        A_t = 0,\qquad 
        \partial_x A_\varphi = 0,
    \end{aligned}
\end{equation}
for $x=-1$, where we have used the gauge freedom of the electromagnetic field to impose the condition on $A_t$. To avoid conical singularities we also impose $h=1$ on the symmetry axis, as explained in Ref. \cite{Fernandes:2022gde}.

Asymptotically we have the fall-offs
\begin{equation}
    g_{tt} \sim -1 + \frac{2M}{r} + \ldots, \quad g_{\varphi t} \sim \frac{2J}{r} \sin^2 \theta, \quad a \sim \frac{D \cos \theta}{r^2}, \quad A_t \sim \Phi - \frac{Q}{r}, \quad A_\varphi \sim \frac{\mu_M}{r},
    \label{eq:asymptotic}
\end{equation}
where $M$, $J$ and $Q$ were defined above, $D$ is the dipole moment of the axion, $\Phi$ is the electrostatic potential difference between infinity and the horizon, and $\mu_M$ is the magnetic dipole moment. The gyromagnetic ratio $\mathtt{g}$ (also known as \emph{g-factor}) measures how the magnetic dipole moment is induced by the total angular momentum and charge for a given mass
\begin{equation}
    \mu_M = \mathtt{g} \frac{Q J}{2M}.
    \label{eq:gyromagnetic}
\end{equation}
For a Kerr-Newman black hole, $\mathtt{g}$ is always equal to $2$. The Hawking temperature $T_H$ \cite{Hawking:1975vcx} and event horizon area $A_H$ are given by
\begin{equation}
  T_H = \left. \frac{1}{2\pi r_H} \frac{f}{\sqrt{g h}} \right|_{x=-1}, \quad A_H = \left. 2\pi r_H^2 \int_{0}^{\pi} d\theta \sin \theta \frac{g \sqrt{h}}{f} \right|_{x=-1},
\end{equation}
with the entropy of the black hole given by $S = A_H/4$. The angular velocity of the horizon is
\begin{equation}
    \Omega_H = \left. W/r_H \right|_{x=-1}.
\end{equation}
To check the accuracy of our solutions we use a Smarr-type relation \cite{PhysRevLett.30.71,Liberati:2015xcp} that black hole solutions should obey
\begin{equation}
    M = 2T_H S + 2 \Omega_H J + \Phi Q - \frac{m_a^2}{4\pi} \int d^3x \sqrt{-g} a^2,
    \label{eq:smarr}
\end{equation}
which relates global charges at infinity and horizon quantities.

To solve the partial differential equations resulting from the field equations, we will use the code described in Ref. \cite{Fernandes:2022gde}, which employs a pseudospectral method together with a Newton-Raphson root-finding algorithm to solve the non-linear system. We expand each of the functions in a spectral series with resolution $N_x$ and $N_\theta$ in the radial and angular coordinates $x$ and $\theta$, respectively. The spectral series we use for each of the functions ${f,g,h,W,A_t,A_\varphi }$ (collectively denoted by $\mathcal{F}$) is given by
\begin{equation}
  \mathcal{F}^{(k)} = \sum_{i=0}^{N_x-1} \sum_{j=0}^{N_\theta-1} \alpha_{ij}^{(k)} T_i(x) \cos \left(2j\theta\right),
\label{eq:spectralexpansion1}
\end{equation}
where $T_i(x)$ denotes the $i^{th}$ Chebyshev polynomial. For the axion, due to its odd parity properties, we use 
\begin{equation}
  a = \sum_{i=0}^{N_x-1} \sum_{j=0}^{N_\theta-1} \alpha_{ij}^{(a)} T_i(x) \cos \left([2j+1]\theta\right).
\label{eq:spectralexpansion2}
\end{equation}
With these spectral expansions, all the angular boundary conditions are automatically satisfied and need not be imposed explicitly in the numerical method.

In our setup, we have five input parameters: $(r_H,\chi, q, \g, m_a)$ and typically use a resolution of $N_x \times N_\theta = 40 \times 8$. For the vast majority of solutions we fix $r_H = 1$ in the code. The error on our solutions is typically less than $\mathcal{O}\left(10^{-12}\right)$, as estimated by the Smarr-type relation \eqref{eq:smarr}, although errors increase for large charges and/or spins. As an initial guess for the solver, we usually use a comparable Kerr-Newman black hole or a previously obtained and similar black hole with axion hair. We focus on studying fundamental solutions (solutions without nodes in the scalar field radial and/or angular profiles), as these are expected to be the (most) stable solutions. 

\section{Results}
\label{sec:results}
\subsection{Small values of spin}
\label{sec:smallspin}
Ref. \cite{Boskovic:2018lkj} studied the stability of Reissner-Nordström black holes for the theory \eqref{eq:action} and concluded that for small electric charge ($q\ll 1$) and axion mass ($M m_a \ll 1$), the system is unstable against axial perturbations with multipole $l$ if $q \gtrsim q^{\mathrm{crit}}$, where
\begin{equation}
  q^{\mathrm{crit}} \approx \frac{2}{\g}\left(1.45 + l + (M m_a)^{3/2} \right).
  \label{eq:couplingcrit}
\end{equation}
We therefore expect at least two branches of solutions to exist: one that exists for any value of $q$ (labeled as branch 1) and another branch that exists for values $q \gtrsim q^{\mathrm{crit}}$ (labeled as branch 2)\footnote{Equivalently, for a fixed dimensionless charge $q$, we expect a new branch of solutions to exist for couplings above $\g^{\mathrm{crit}} \sim \frac{2}{q}\left(1.45 + l + (M m_a)^{3/2} \right)$.}.

\begin{figure}[]
  \centering
    \includegraphics[width=.5\textwidth]{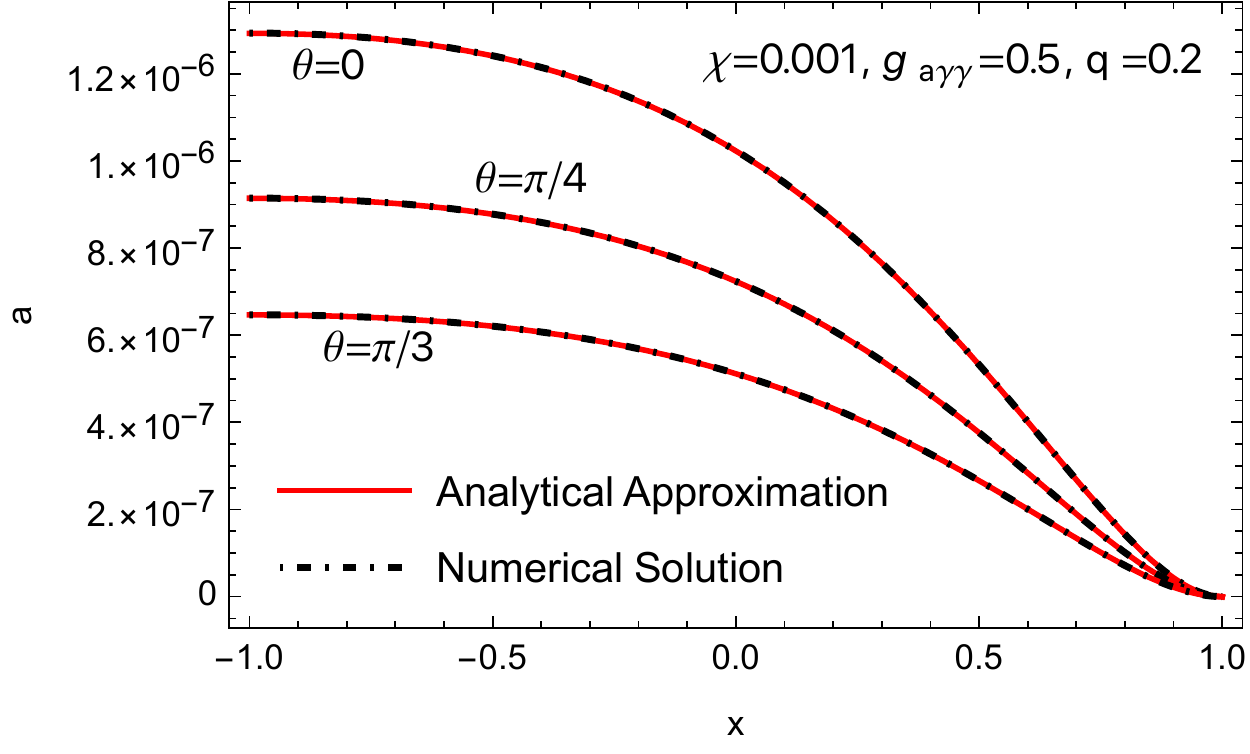}
    \caption{Comparison between the perturbative solution in Eq. \eqref{eq:axionpert} and our numerical results for the axion for a solution with $\chi = 0.001$, $\g = 0.5$, and $q=0.2$. The initial guess provided to the code to obtain this solution was a Kerr-Newman black hole with the same $\chi$ and $q$, and a vanishing axion profile.}
  \label{fig:compare_pert}
\end{figure}

The first branch's solutions for a massless axion were also constructed perturbatively in Ref. \cite{Boskovic:2018lkj} to first order in spin $\chi$ and axion coupling $\g$ around a Kerr-Newman black hole. To this order, non-trivial corrections appear only at the level of the axion and they read
\begin{equation}
    a \approx \g M \chi \left[ \frac{2}{r_{BL}^+} - \frac{1}{r_{BL}} + \frac{2r_{BL} - r_{BL}^+ - r_{BL}^-}{r_{BL}^+ r_{BL}^-} \log\left(1-\frac{r_{BL}^-}{r_{BL}}\right) \right] \cos \theta,
    \label{eq:axionpert}
\end{equation}
where $r_{BL}$ is the Boyer-Lindquist radial coordinate, related to the one in our coordinate system via Eq. \eqref{eq:rbl}, and $r_{BL}^{\pm} = M\left(1\pm \sqrt{1 - q^2 - \chi^2}\right)$.
\begin{figure}[]
  \centering
    \includegraphics[width=.5\textwidth]{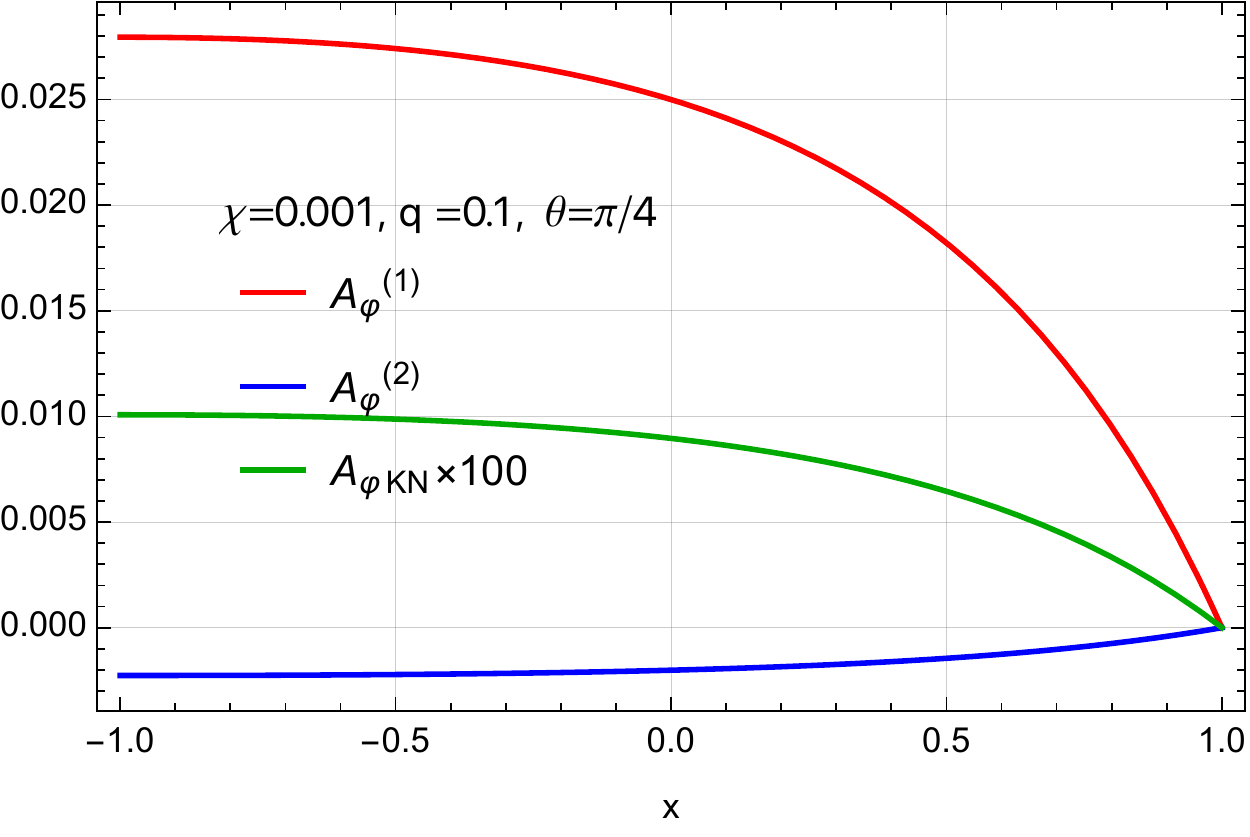}\hfill
    \includegraphics[width=.5\textwidth]{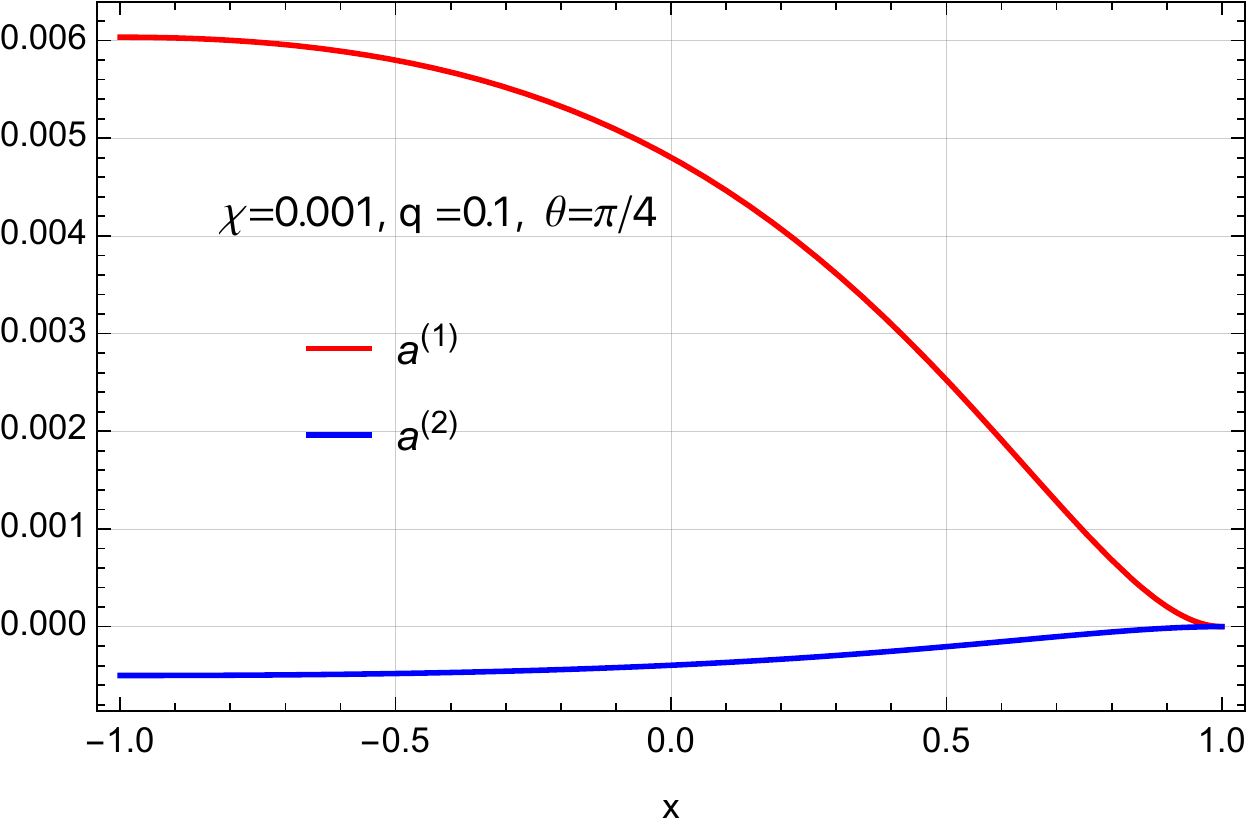}\vfill
    \caption{The radial profiles of the function $A_\varphi$ and field $a$ are shown for two solutions with $\g=50$, $\chi=0.001$, and $q=0.1$, both plotted for an illustrative value of $\theta=\pi/4$. The solution belonging to branch 1 is labeled (1), while the solution belonging to branch 2 is labeled (2). The profiles of $A_\varphi$ and $a$ for the two solutions are markedly different.}
  \label{fig:2solutions}
\end{figure}
In this section, our focus is on constructing non-perturbative solutions and examining the existence of two solution branches. Specifically, we will concentrate on the small spin limit, where we take $\chi = 0.001$ to ensure that Eq. \eqref{eq:couplingcrit} holds. As another test to the code we have compared our numerical results with the perturbative solution in Eq. \eqref{eq:axionpert}, observing great agreement as shown in Fig. \ref{fig:compare_pert}.

By taking $q=0.1$, we can observe the presence of the two solution branches. The first branch exists for all values of $\g$, while the second branch is only attainable for values of $\g \gtrsim 49$, in agreement with Eq. \eqref{eq:couplingcrit}. In all of the numerical solutions we obtained, we found that the metric function $g(r,\theta)$ matched the analytical results from the previous section with high accuracy. However, the function $A_\varphi$, which is related to the magnetic field of the system, is significantly amplified compared to a comparable Kerr-Newman black hole. The two branches of solutions exhibit drastic differences in the profiles of $A_\varphi$ and $a$, having different signs. This has implications for the gyromagnetic ratio of the solutions defined in Eq. \eqref{eq:gyromagnetic}. An example of this can be seen in Fig. \ref{fig:2solutions}, where the branch 1 solution (represented by the red line) has $\mathtt{g} \approx 575.02$ and the branch 2 solution (represented by the blue line) has a negative gyromagnetic ratio of $\mathtt{g} \approx -44.93$ (recall that for a Kerr-Newman black hole, $\mathtt{g}=2$ always).

To gain a deeper understanding of the solution branches and their properties, we traced each branch through its domain of existence while monitoring various properties, including the gyromagnetic ratio, entropy, Hawking temperature, and the Gauss-Bonnet curvature scalar on the horizon. Specifically, we looked at the $\g=50$ case with $\chi=0.001$, and varied the parameter $q$. Based on Eq. \eqref{eq:couplingcrit}, we anticipated that the second branch of solutions would emerge for values $q \gtrsim 0.098$.
\begin{figure}[]
  \centering
    \includegraphics[width=.5\textwidth]{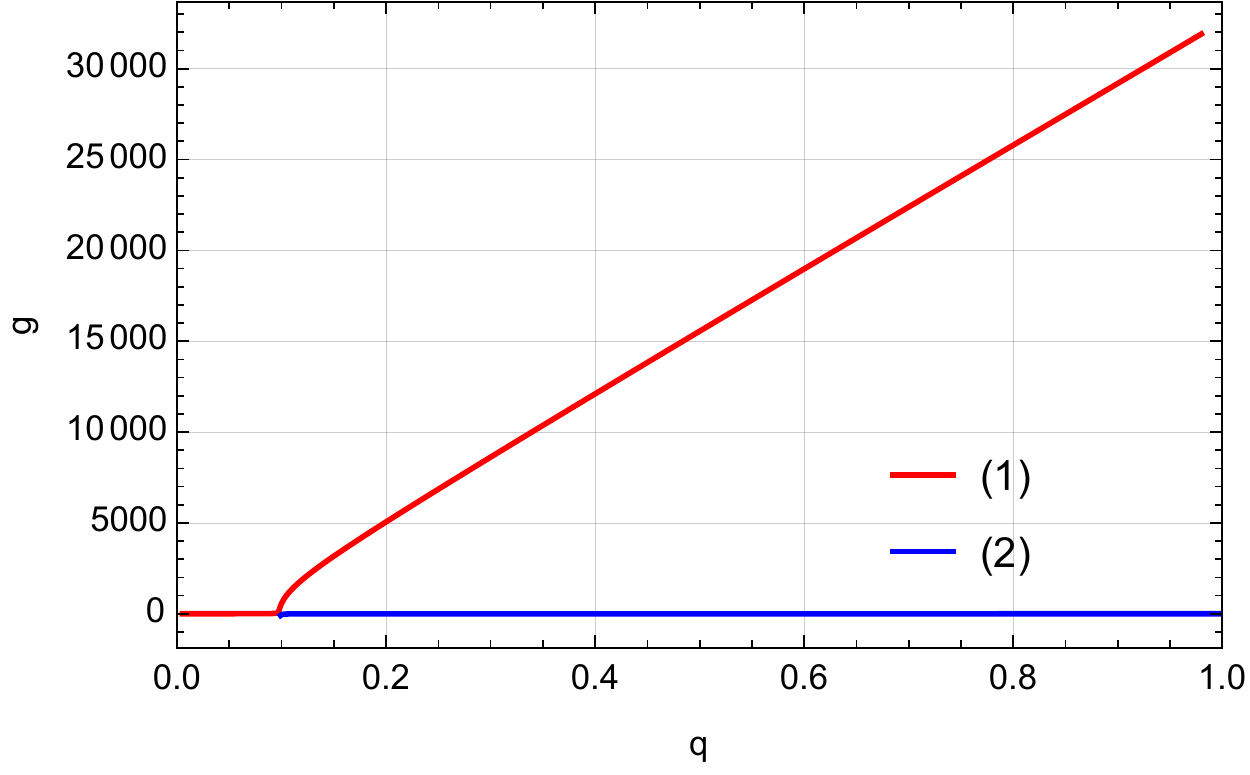}\hfill
    \includegraphics[width=.5\textwidth]{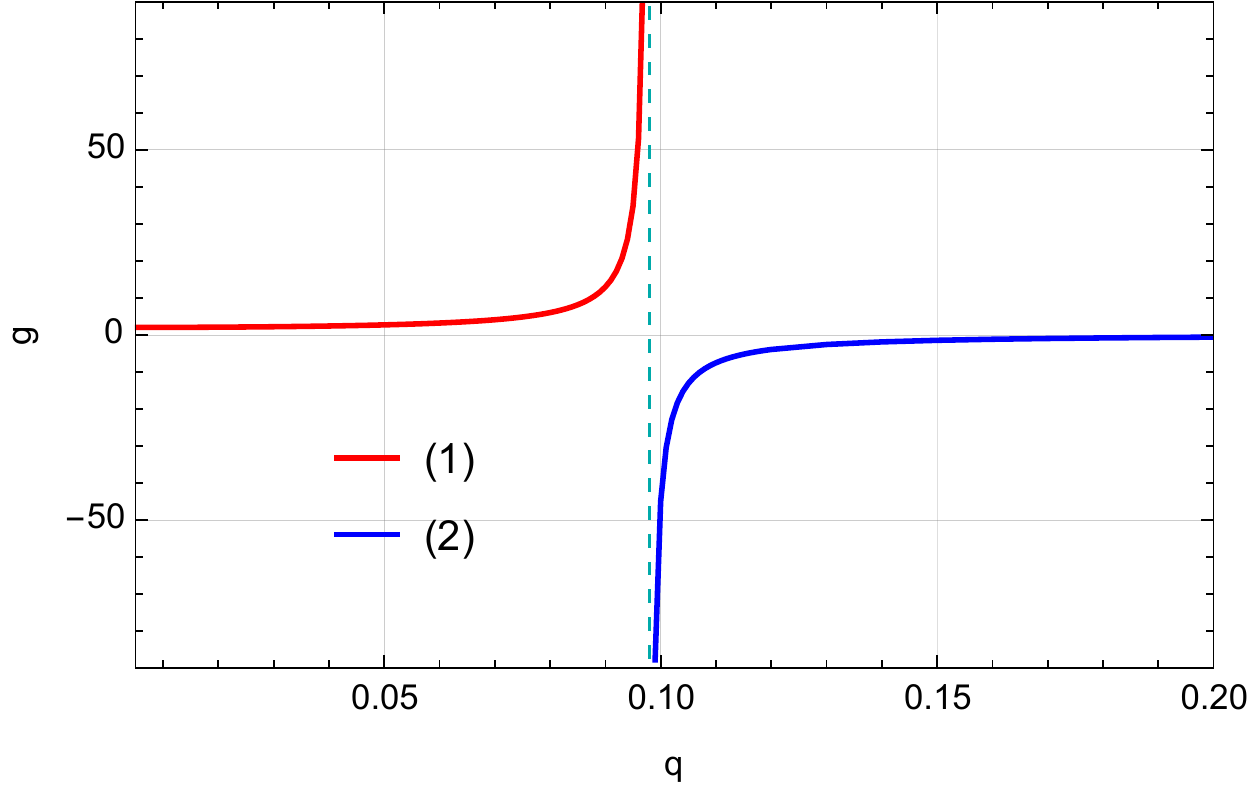}\vfill
    \caption{The figure shows the behavior of the gyromagnetic ratio $\mathtt{g}$ for the two branches of solutions. The left panel displays a sharp increase in the gyromagnetic ratio for first branch solutions around $q \approx 0.098$, coinciding with the appearance of a second branch of solutions. A more detailed view of this feature is provided in the right panel, which is a zoomed-in version of the plot on the left. These solutions were obtained with $\chi=0.001$ and $\g=50$.}
  \label{fig:gyromagnetic}
\end{figure}
Fig. \ref{fig:gyromagnetic} displays the results for the gyromagnetic ratio, where the emergence of the second branch of solutions is manifest. We observe that the first branch starts at values close to $\mathtt{g} \approx 2$ for small values of $q$, similar to a Kerr-Newman black hole. However, once $q$ increases, we see a significant departure from this behavior, with a sharp increase occurring at around $q \approx 0.098$ when the second branch of solutions emerges. This second branch of solutions has negative values of $\mathtt{g}$, with very large magnitudes observed for $q \approx 0.098$, and drives the gyromagnetic ratio to small values as $q$ becomes larger, until the branch of solutions ends. However, we have noticed some features in the gyromagnetic ratio of the second branch for certain discrete values of the charge $q$. Although these features are too small to be visible in a plot for the small spin case, we will discuss them in more detail in the next section for higher spins.

\begin{figure}[]
  \centering
    \includegraphics[width=.5\textwidth]{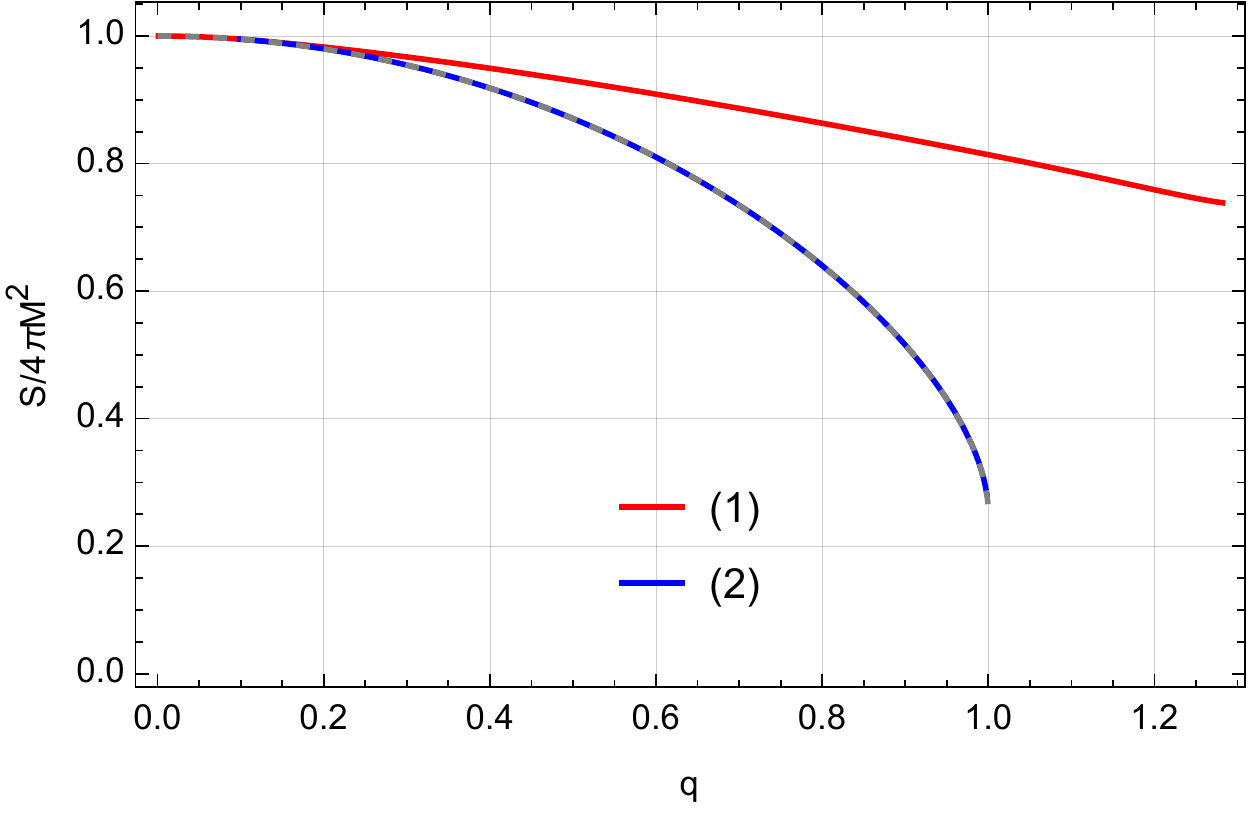}\hfill
    \includegraphics[width=.5\textwidth]{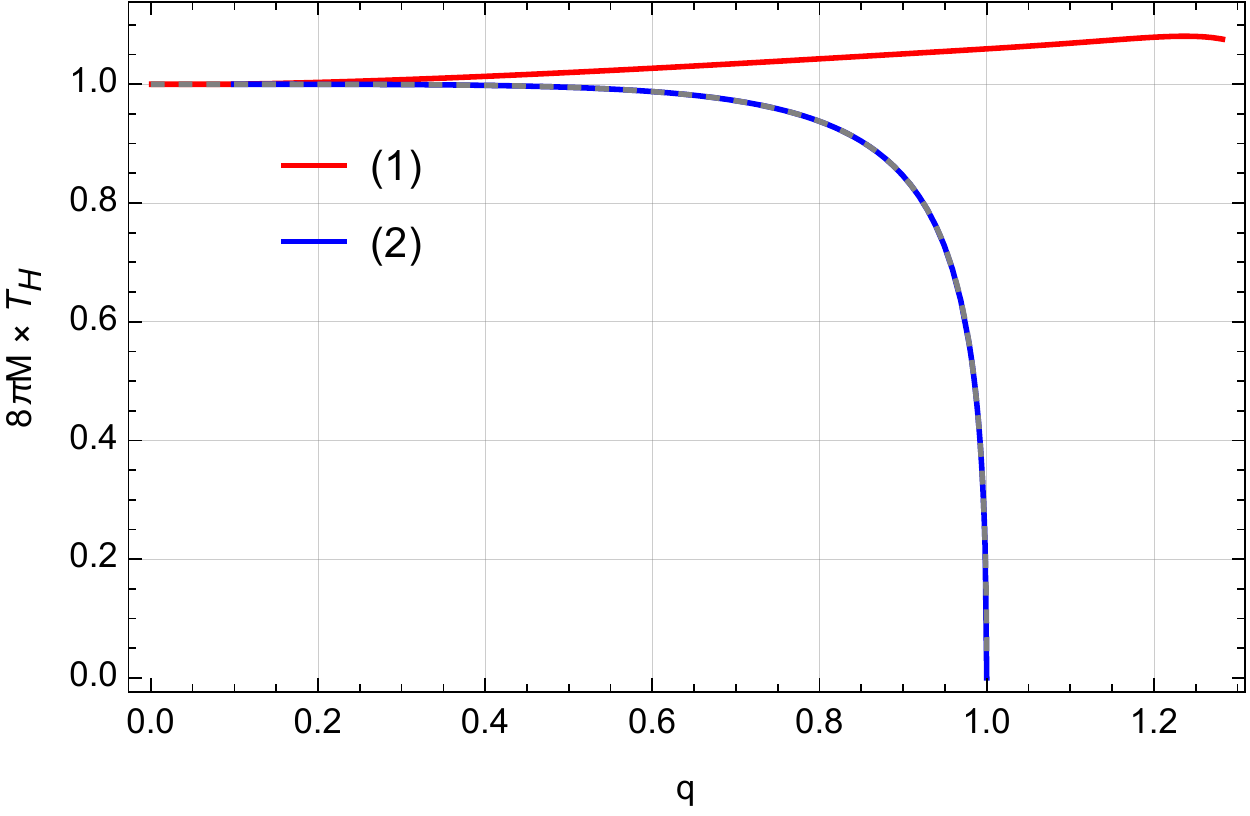}\vfill
    \caption{The left plot shows the behavior of the entropy and the right plot shows the behavior of the Hawking temperature for both branches of solutions, with a dashed gray line indicating the profile for a Kerr-Newman black hole. The values of $\chi=0.001$ and $\g=50$ were used to obtain these solutions.}
  \label{fig:entropy_T}
\end{figure}

\begin{figure}[]
  \centering
    \includegraphics[width=.5\textwidth]{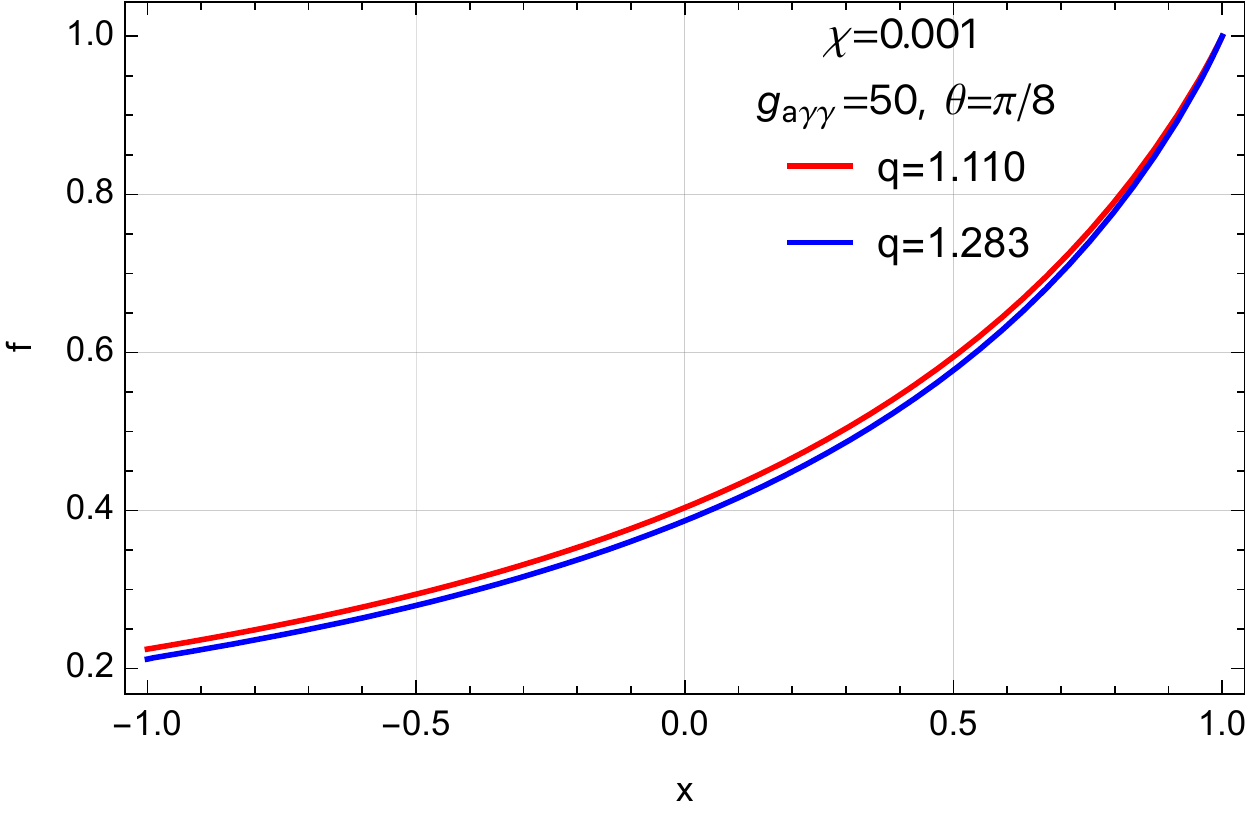}\hfill
    \includegraphics[width=.5\textwidth]{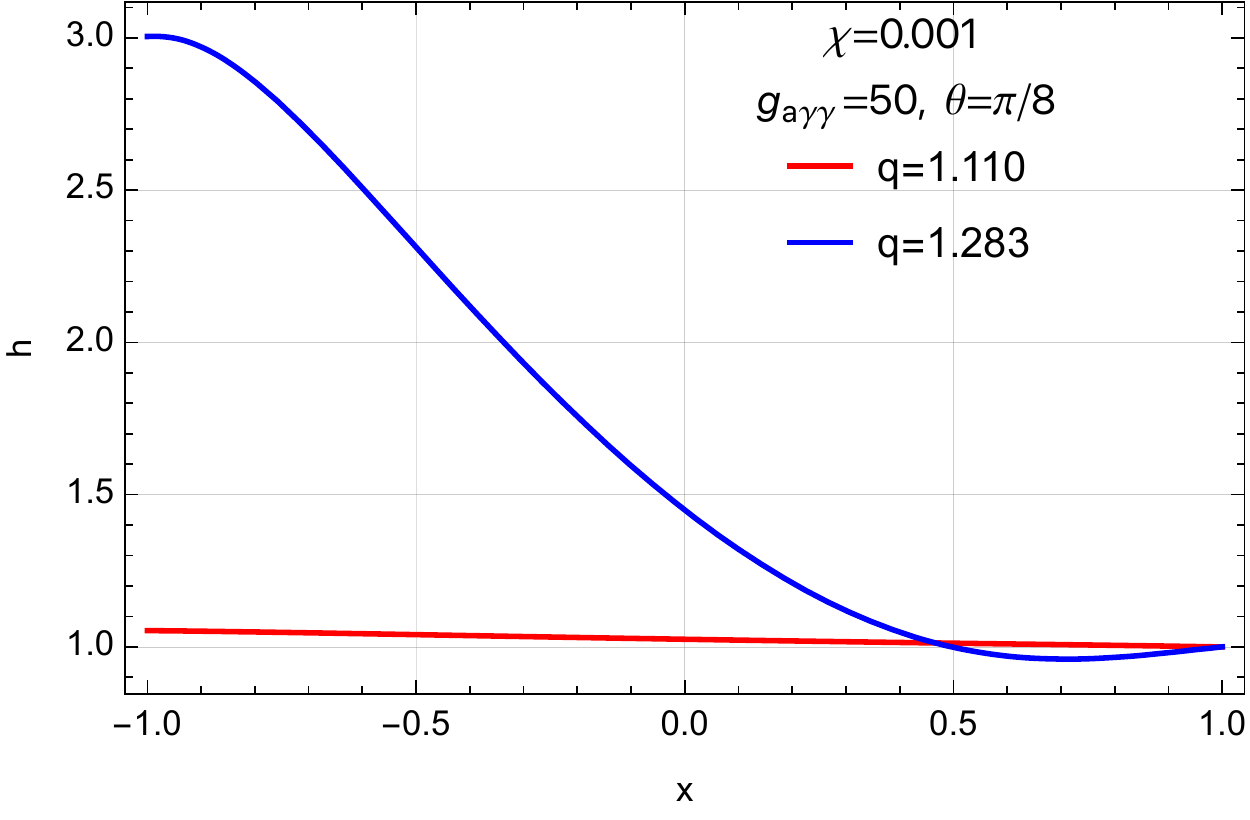}\vfill
    \includegraphics[width=.5\textwidth]{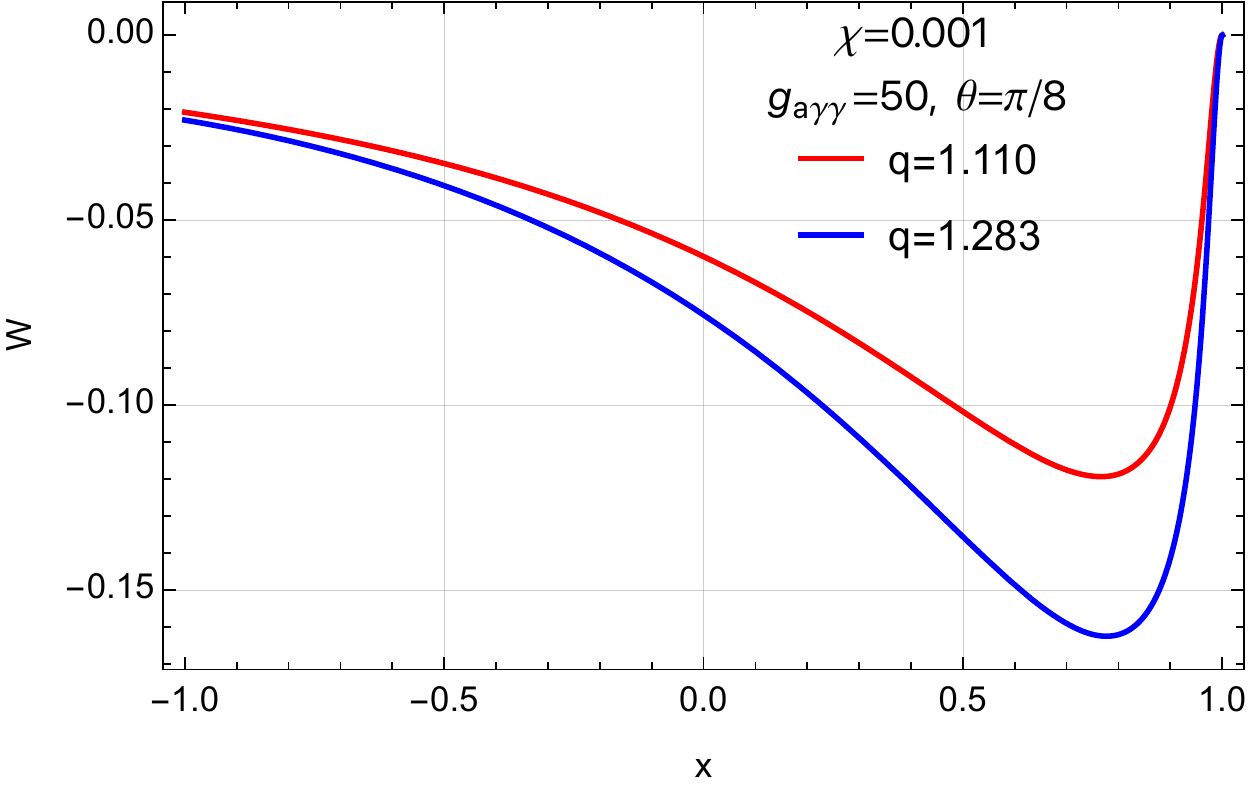}\hfill
    \includegraphics[width=.5\textwidth]{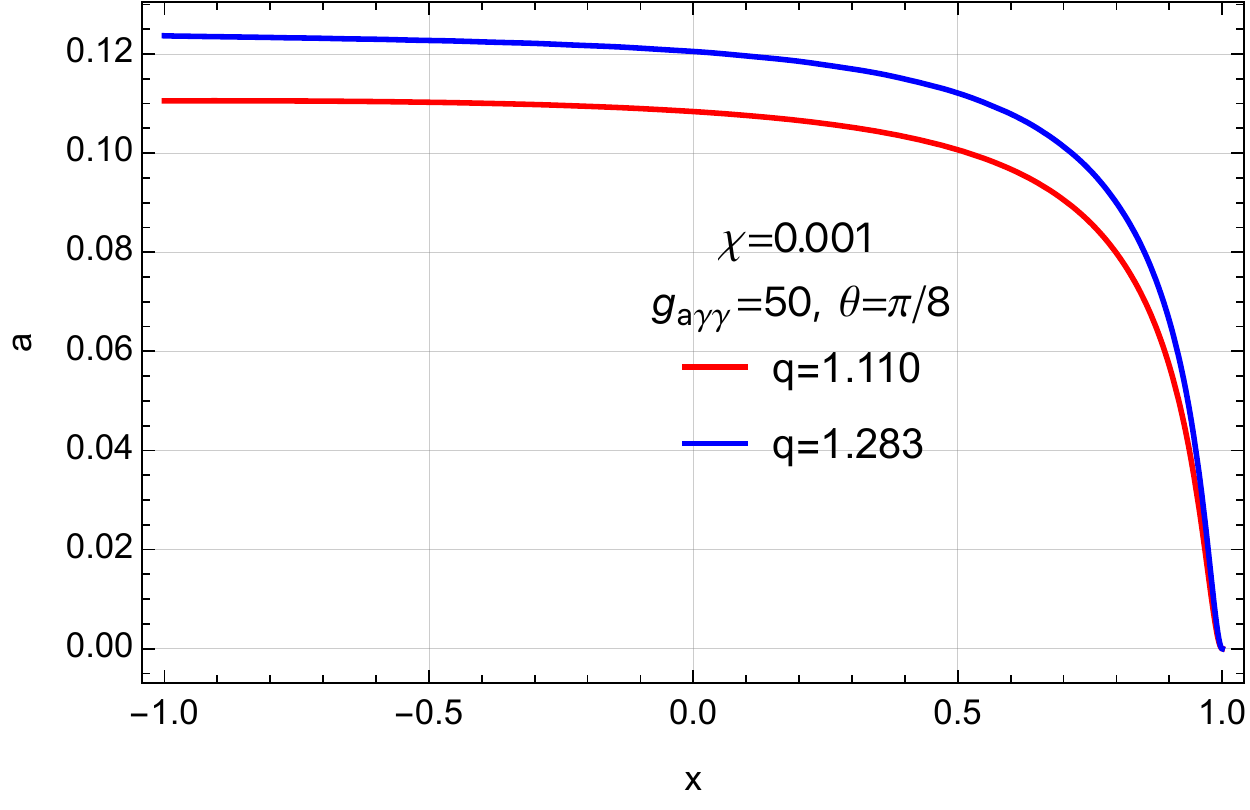}\vfill
    \includegraphics[width=.5\textwidth]{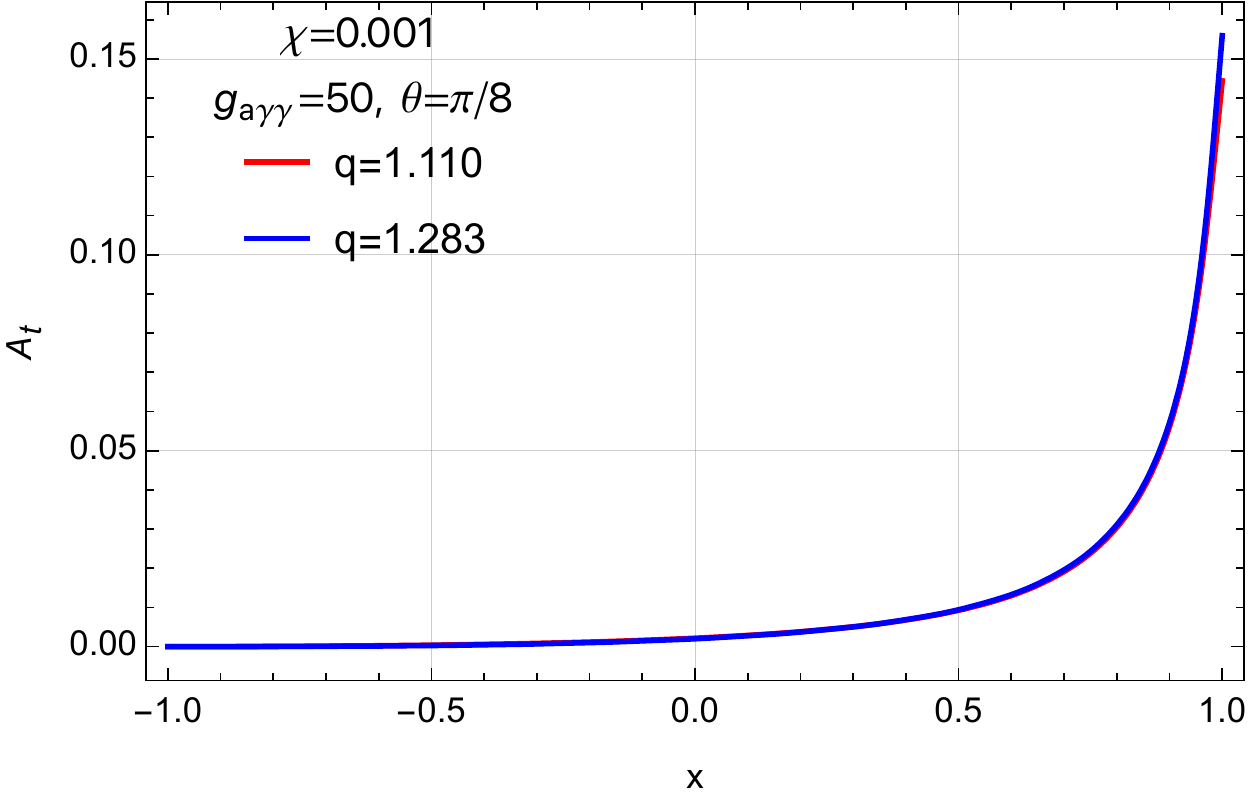}\hfill
    \includegraphics[width=.5\textwidth]{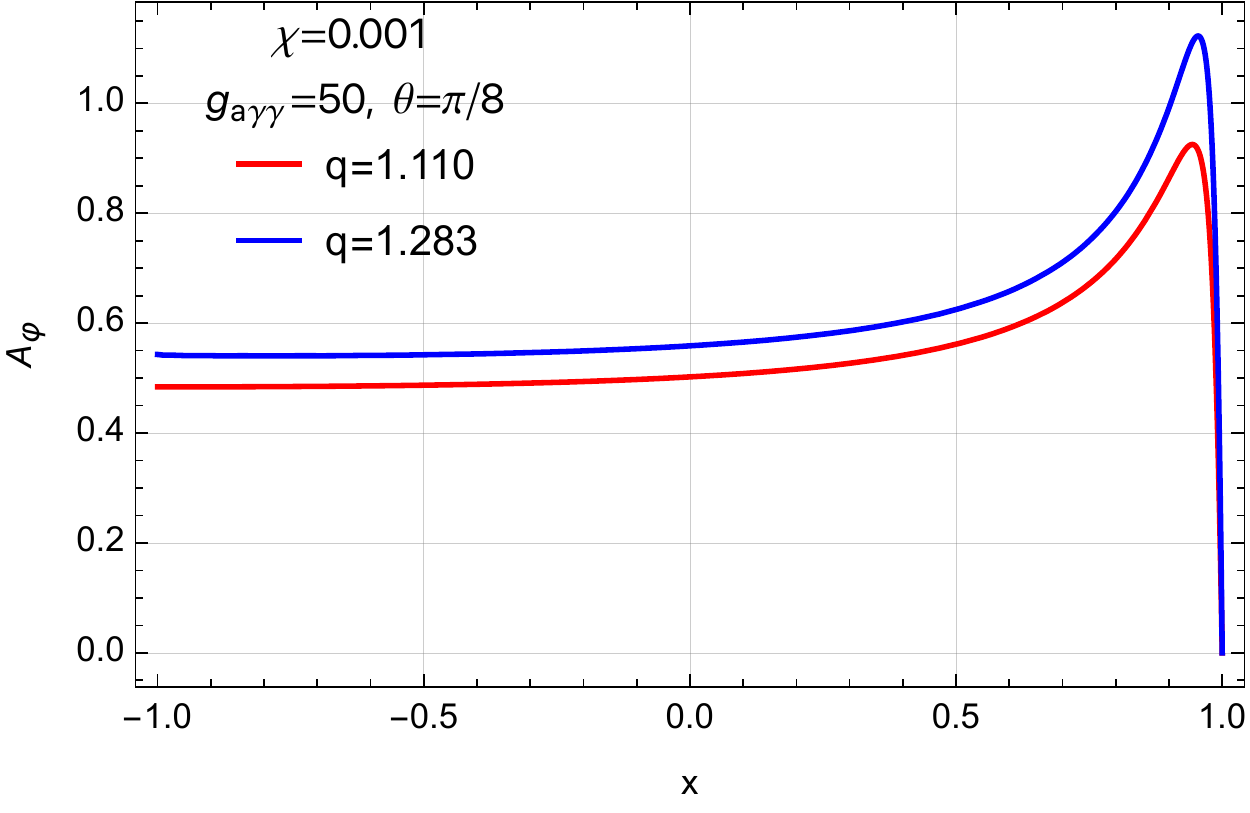}\vfill
    \caption{Radial profiles of two solutions belonging to the first branch that violate the Kerr-Newman bound, one of which is in close proximity to the singular solution. The value of $\theta$ is fixed at $\pi/8$.}
  \label{fig:overcharged}
\end{figure}

In Figure \ref{fig:entropy_T}, we can see the plots of entropy and Hawking temperature for solutions with small spin $\chi=0.001$. It is observed that the second branch of solutions behaves similarly to a Kerr-Newman black hole, whereas the first branch differs significantly. In particular, the first branch does not approach an extremal black hole as we increase $q$, because the Hawking temperature does not approach zero.
Instead, the numerical monitoring of the Gauss-Bonnet curvature scalar on the horizon suggests the presence of a curvature singularity on the horizon as we approach the maximum value of $q$.
Additionally, in the domain where both branches of solutions co-exist, the first branch is always entropically preferred over a comparable second branch solution. Moreover, it is possible for first branch solutions to violate the Kerr-Newman bound, which states that $q$ should be less than or equal to $\sqrt{1-\chi^2}$, which, for small spins, is approximately $q\lesssim 1$. Fig. \ref{fig:overcharged} presents the radial profiles of the involved functions and fields for two solutions belonging to the first branch that violate the Kerr-Newman bound, one of which is very close to the singular solution. It should be noted that while the analytical examples presented in Ref. \cite{Astefanesei:2019pfq} and Ref. \cite{Narang:2020bgo} for related theories feature endpoint solutions that are singular, those solutions have a vanishing horizon area, which is not observed in our solutions.

\begin{figure}[]
  \centering
    \includegraphics[width=.5\textwidth]{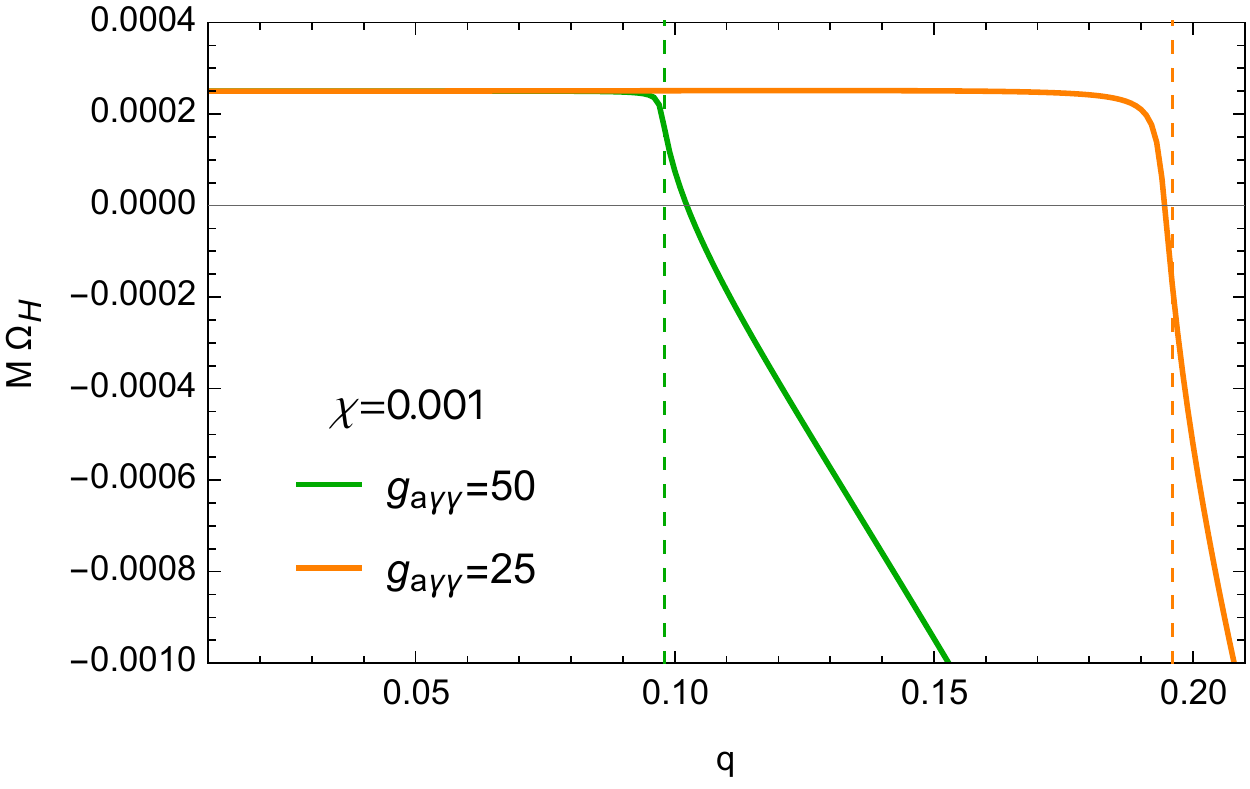}\hfill
    \caption{Angular velocity of the horizon as a function of the charge to mass ratio $q$ for solutions from the first branch. The vertical dashed lines indicate the critical values $q_{\mathrm{crit}}$ obtained using Eq. \eqref{eq:couplingcrit} for each coupling constant. It is evident that the horizon angular velocity undergoes a transition from positive to negative values around $q=q_{\mathrm{crit}}$.}
  \label{fig:counter_rot}
\end{figure}

We have observed another intriguing characteristic of the first branch solutions, namely the existence of counterrotating black holes where the horizon angular velocity $\Omega_H$ and the total angular momentum $J$ have opposite signs. This can be seen in Fig. \ref{fig:counter_rot}, where the black hole solutions initially have a positive $\Omega_H$ for small values of $q$, but undergo a transition from positive to negative values of $\Omega_H$ at around $q=q^{\mathrm{crit}}$. This feature is also observed in the solutions of Fig. \ref{fig:overcharged}. Counterrotating black holes have been previously noted in works such as Refs. \cite{Kunz:2006xk, Kleihaus:2003df, Kunz:2005ei}.

\begin{figure}[]
  \centering
    \includegraphics[width=.5\textwidth]{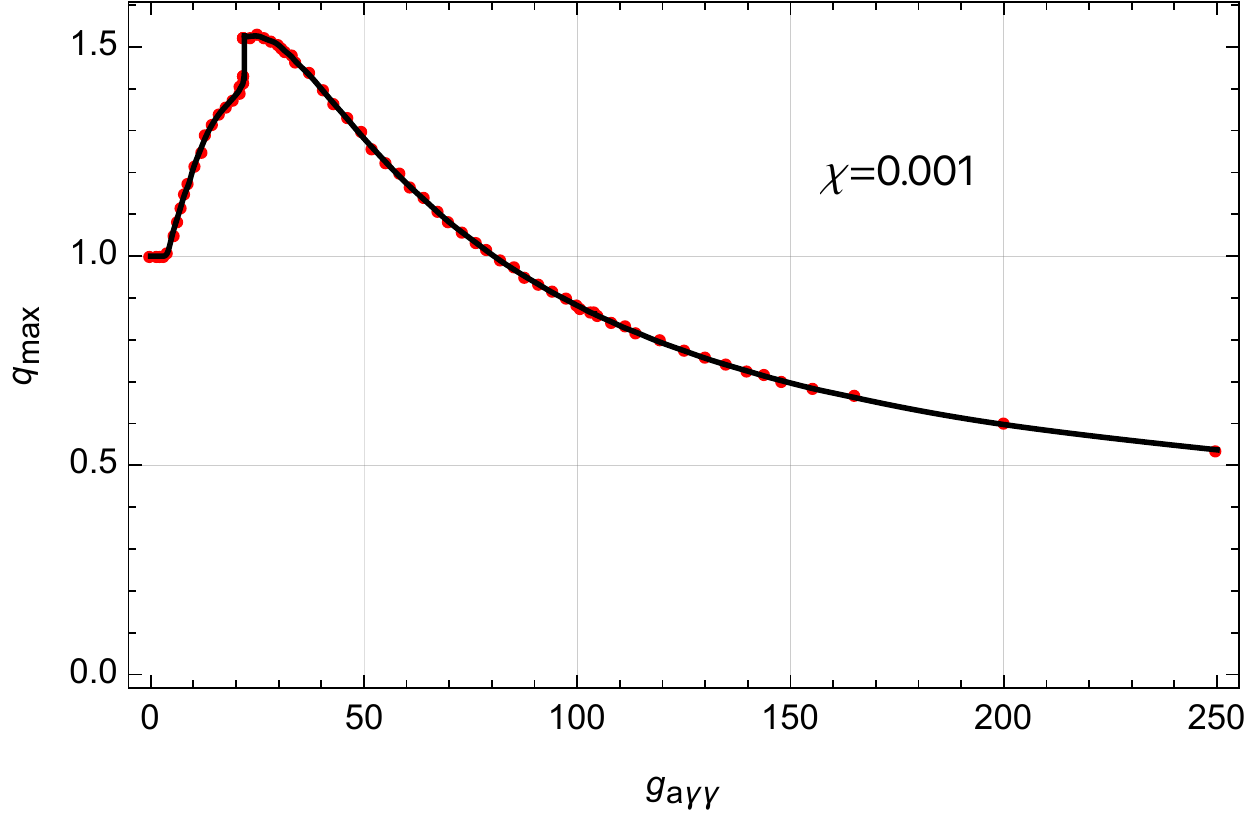}
    \caption{The plot shows the maximum charge $q_{max}$ allowed by the first branch solutions as a function of the coupling constant $\g$, with a fixed spin value of $\chi=0.001$. The maximum value of $q_{max}$ is approximately 1.525, which is obtained for a value of $\g$ around 25. It can be observed that $q_{max}$ experiences a significant growth around $\g \approx 22$.}
  \label{fig:qmax}
\end{figure}

Fig. \ref{fig:qmax} shows that the maximum charge $q_{max}$ allowed by the first branch solutions strongly depends on the coupling $\g$. Specifically, we observe that the violations of the Kerr-Newman bound are only possible for values of $\g$ less than or around $80$, for a fixed spin value of $\chi = 0.001$. For higher values of the coupling, charge to mass ratios are always less than unity.

\subsection{Higher spins}
\label{sec:highspin}
\begin{figure}[]
  \centering
    \includegraphics[width=.5\textwidth]{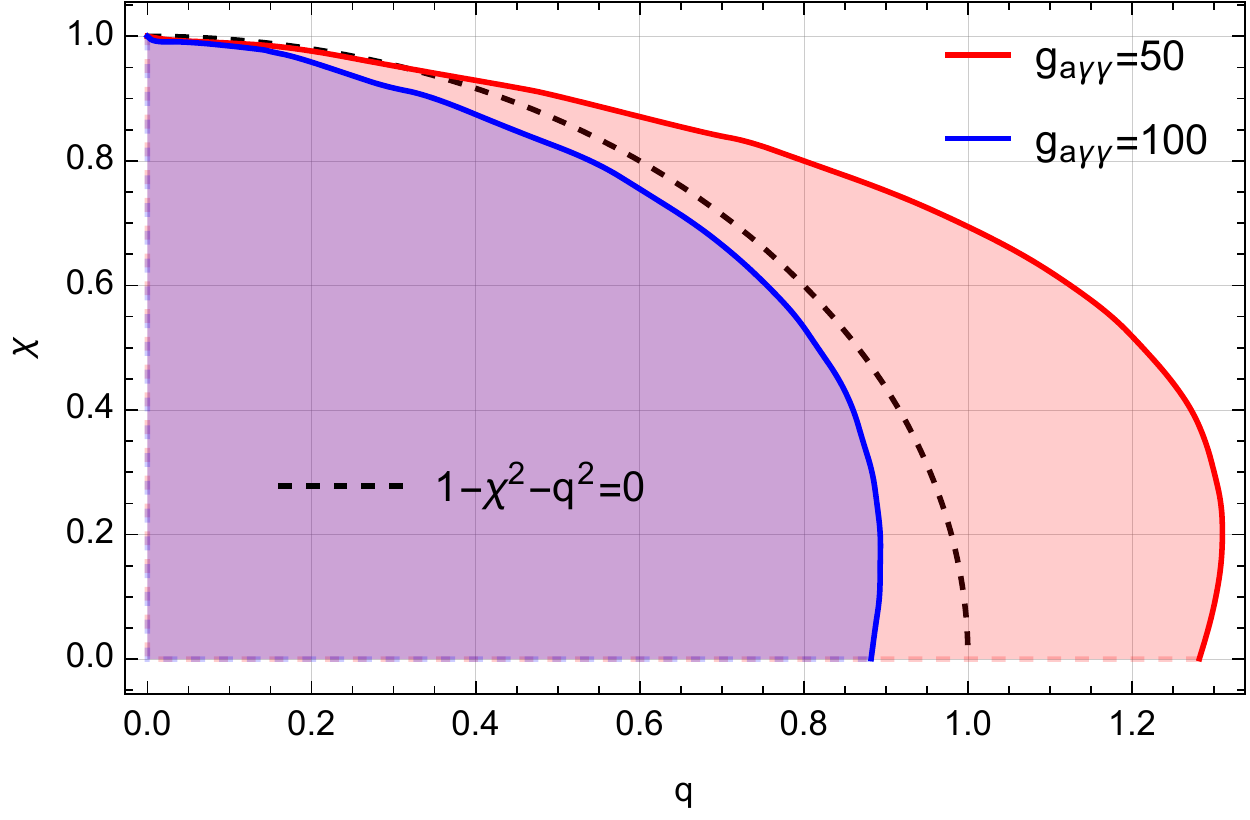}\hfill
    \includegraphics[width=.5\textwidth]{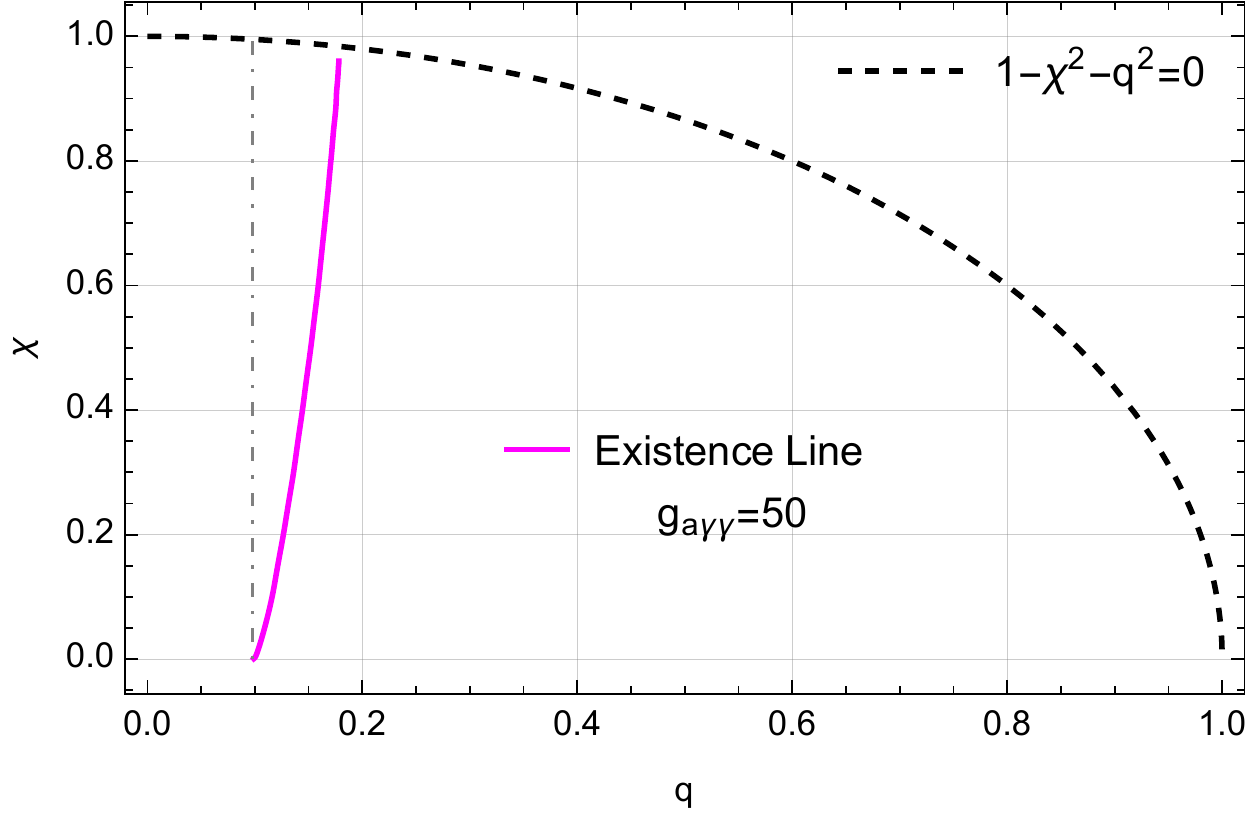}
    \caption{(Left) The region in parameter space where solutions of the first branch exist is shown for $\g=50$ and $\g=100$. (Right) Solutions outside the first branch exist for values of the charge to mass ratio $q$ greater than those shown in the pink line for a coupling $\g=50$. The dot-dashed gray line indicates the value predicted by Eq. \eqref{eq:couplingcrit} for small spin, showing excellent agreement.}
  \label{fig:domain50}
\end{figure}

In this section, we examine axionic solutions with higher spin values $\chi$. The full domain of existence for first branch solutions with $\g=50$ ($\g=100$) is presented in Fig. \ref{fig:domain50} on the left, shaded in red (blue), respectively. These solutions can violate the Kerr-Newman bound (dashed line) throughout their domain, with values as high as $q \approx 1.31$ observed for spins $\chi \approx 0.2$ in the $\g=50$ case. Approaching the red and blue lines, we observed a divergent behavior of the Gauss-Bonnet scalar on the horizon. As in the case of small spins, the maximum allowed value of charge $q$ depends heavily on the coupling $\g$, as demonstrated by comparing the $\g=50$ and $\g=100$ cases.
In a manner similar to the case of small spins, we have also observed the presence of solutions that do not belong to the first branch. As we discussed earlier, for a fixed value of coupling, these solutions exist only above a critical value of the charge $q$. While in the case of small spins, we can predict the critical charge value from Eq. \eqref{eq:couplingcrit}, numerical solutions are needed to determine the bifurcation points for higher spins. The existence line for these solutions for $\g=50$ was computed up to spins $\chi \approx 0.96$ and is presented in the right panel of Fig. \ref{fig:domain50}. It is noteworthy that for higher spins, the critical value of the charge increases.
\begin{figure}[]
  \centering
    \includegraphics[width=.5\textwidth]{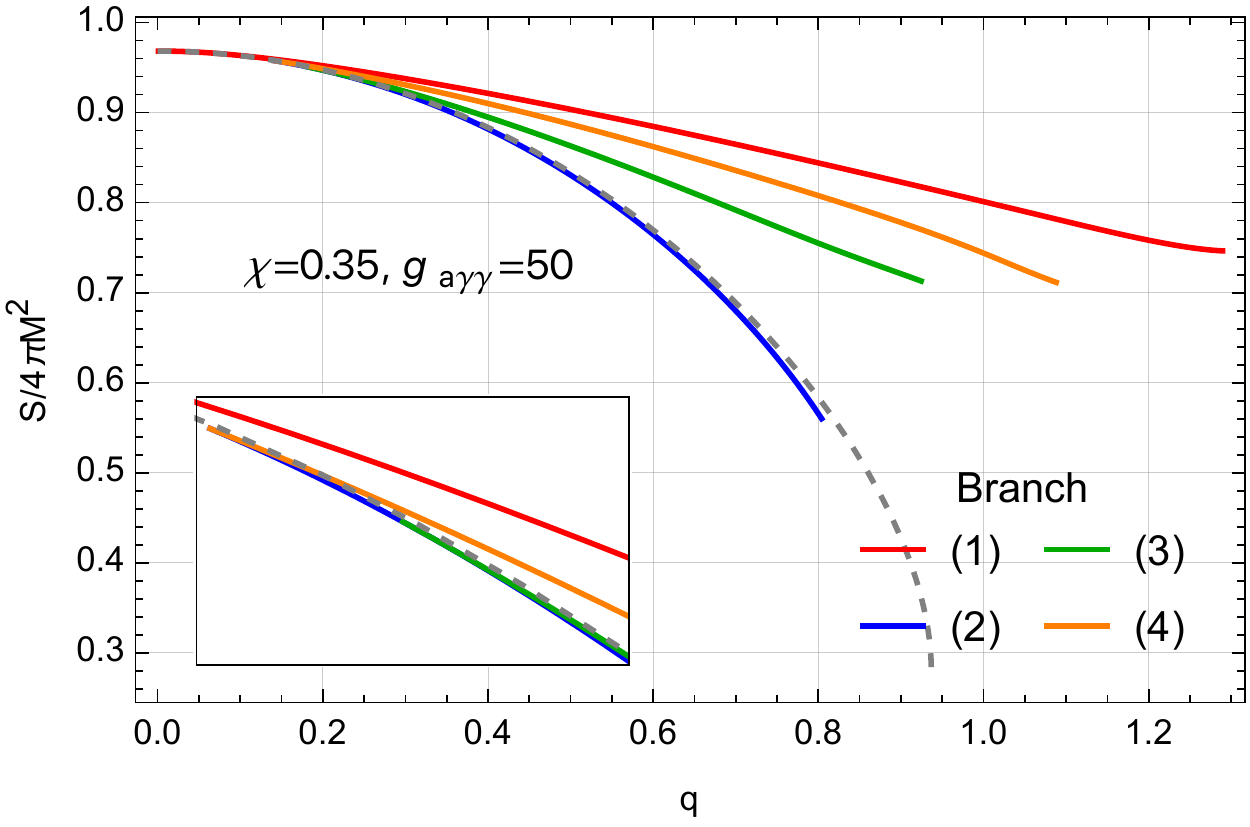}\hfill
    \includegraphics[width=.5\textwidth]{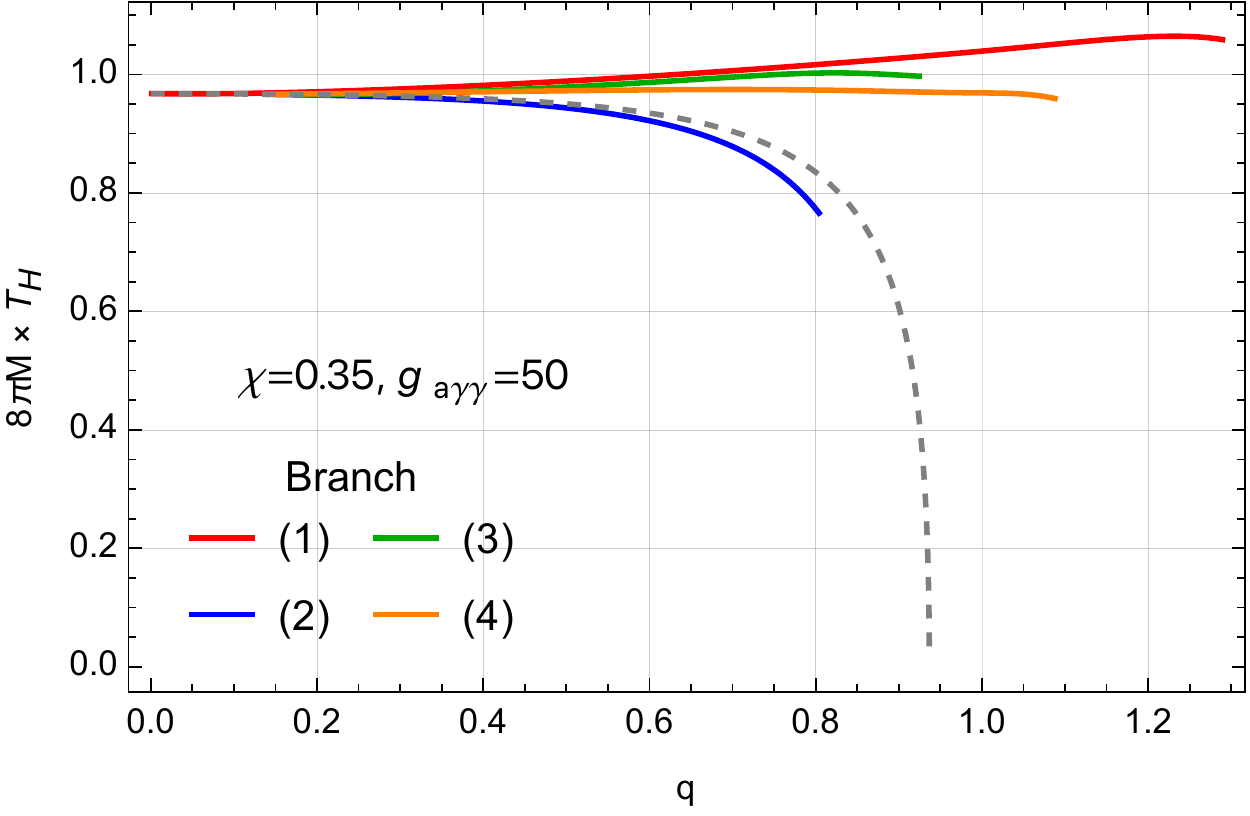}\vfill
    \includegraphics[width=.5\textwidth]{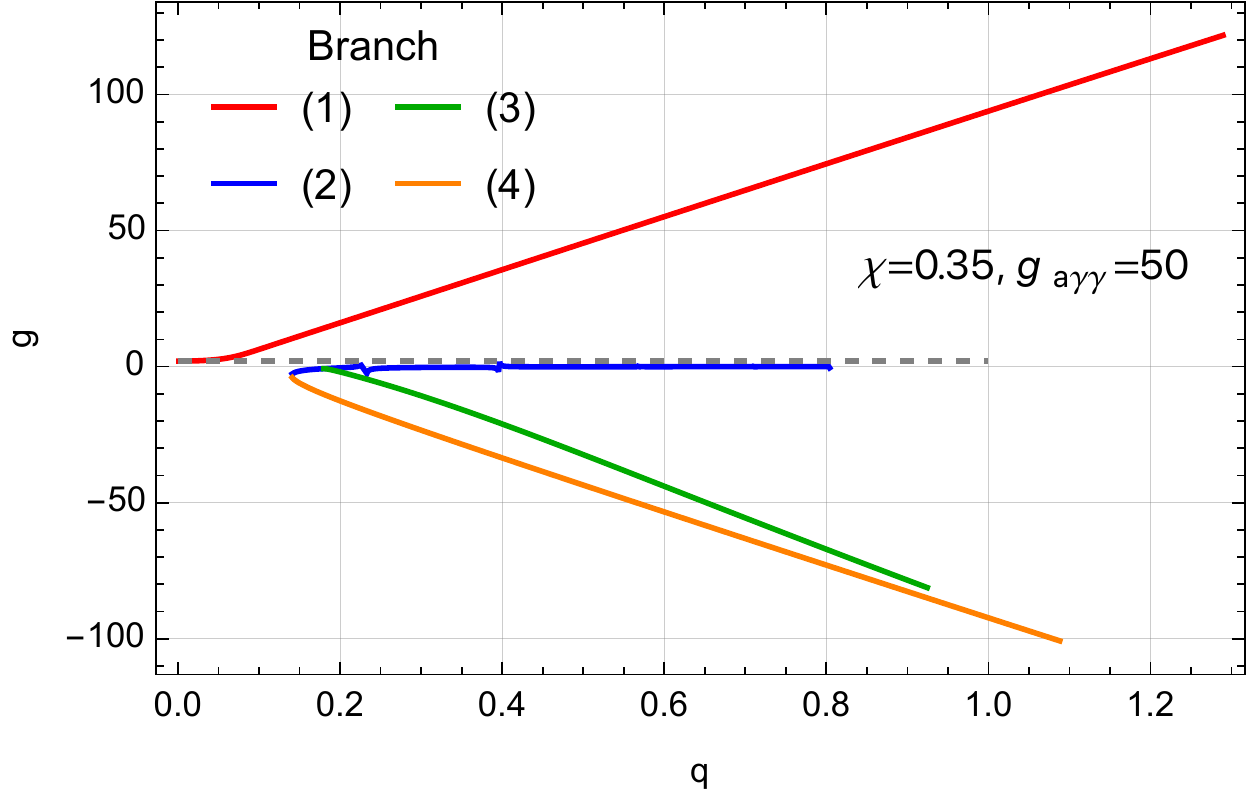}\hfill
    \includegraphics[width=.5\textwidth]{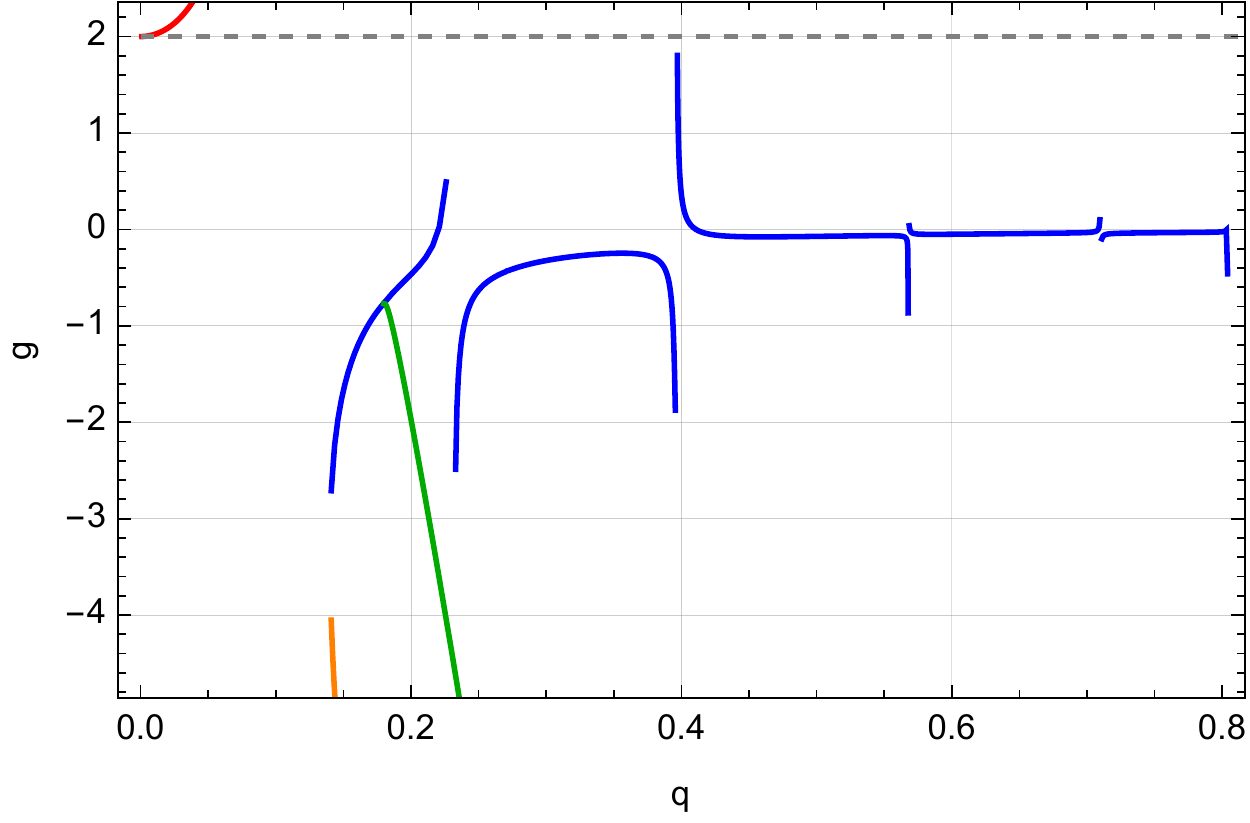}\hfill
    \caption{The behavior of the entropy (top left), Hawking temperature (top right), and gyromagnetic ratio (bottom) are shown for the four solution branches. The plot in the bottom right corner displays a zoomed-in region of the plot in the bottom left corner. The dashed gray line shows the profiles for a Kerr-Newman black hole for reference. The solutions were obtained with $\chi=0.35$ and $\g=50$.}
  \label{fig:branches_035}
\end{figure}
In the case of $\chi=0.35$, at least four distinct branches of solutions were observed, including the first branch, whose domain of existence was presented in Fig. \ref{fig:domain50}, as well as three others with markedly different properties, as shown in Fig. \ref{fig:branches_035}. The second branch corresponds to the second branch seen in the small spin limit, but with the added observation of a resonance phenomenon on the gyromagnetic ratio of the solutions at certain discrete values of the charge $q$. In the specific case shown in Fig. \ref{fig:branches_035}, resonances were observed at $q \approx (0.141, 0.230, 0.396, 0.568, 0.710, 0.804)$. A third new branch was found to bifurcate from the second branch and did not violate the Kerr-Newman bound, while a fourth new branch began at the existence line and did violate it. In the domain of co-existence, for the same charge and spin, the entropy of the solutions in the branches (normalized by $4\pi M^2$) was observed to be ordered as $S_1>S_4>S_3>S_2$. All solutions we studied exhibit ergoregions, light rings, and innermost stable circular orbits that have the same topology as those of a comparable Kerr-Newman black hole.

\section{Conclusions}
\label{sec:conclusions}
In this study, we investigated hairy black hole solutions in General Relativity minimally coupled to electromagnetic and axion fields. When electric charge and rotation is considered, the coupling of the axion to the electromagnetic field results in hairy solutions due to $\dFF \neq 0$, which sources the scalar field equation \eqref{eq:axionEq}. We solved the field equations of the theory \eqref{eq:action} numerically, using a pseudospectral method \cite{Fernandes:2022gde}.

For small spin values, we verified the existence of two branches of solutions predicted in Ref. \cite{Boskovic:2018lkj}. The critical charge value at which a second branch of solutions appears agreed remarkably well with their prediction. The two branches of solutions have opposite signs for the magnetic function $A_\varphi$ and the axion, leading to gyromagnetic ratios with opposite signs. The first branch of solutions was always found to be preferred based on entropic considerations, and violations of the Kerr-Newman bound were observed for these solutions. Additionally, counterrotating black holes, where the angular velocity at the horizon and total angular momentum have opposite signs, were observed.

For higher spin values, we found at least four distinct branches of solutions, all with unique properties. Again, the first branch was preferred based on entropic considerations and was present for all charge/coupling values. We computed the domain of existence for first branch solutions for two distinct coupling values and the existence line at which solutions for other branches start to emerge.

Avenues for future research include a deeper study of the massive axion case, and of phenomenological properties of these solutions, such as ergoregions, light rings, innermost stable circular orbits, and shadows. It would also be interesting to investigate the polarization effects of light rays passing through our axion profiles \cite{Plascencia:2017kca,Chen:2019fsq,Chen:2021lvo}.

Our study provides the first non-perturbative numerical exploration of spinning hairy black hole solutions in the axion model presented in Eq.~\eqref{eq:action}. These solutions exhibit intriguing properties that are different from those of typical Kerr-Newman black holes. Our publicly available code \cite{gitlink} can be used to further investigate these solutions in future studies.

\subsection*{Acknowledgments}
\noindent C.B. and P.F. are supported by a Research Leadership Award from the Leverhulme Trust. C.B. is also supported by the STFC under grant ST/T000732/1. R.B. acknowledges financial support provided by FCT/Portugal, under the Scientific Employment Stimulus -- Individual Call -- 2020.00470.CEECIND. V.C.\ is a Villum Investigator and a DNRF Chair, supported by VILLUM Foundation (grant no.\ VIL37766) and the DNRF Chair program (grant no. DNRF162) by the Danish National Research Foundation. V.C.\ acknowledges financial support provided under the European Union's H2020 ERC Advanced Grant ``Black holes: gravitational engines of discovery'' grant agreement
no.\ Gravitas--101052587. Views and opinions expressed are however those of the author only and do not necessarily reflect those of the European Union or the European Research Council. Neither the European Union nor the granting authority can be held responsible for them.
This project has received funding from the European Union's Horizon 2020 research and innovation programme under the Marie Sklodowska-Curie grant agreement No 101007855.
We acknowledge financial support provided by FCT/Portugal through grants 
2022.01324.PTDC, PTDC/FIS-AST/7002/2020, UIDB/00099/2020 and UIDB/04459/2020.

\appendix
\numberwithin{equation}{section}
\section{The Kerr-Newman Black Hole}
\label{ap:KN}
The Kerr-Newman black hole solution solves the Einstein-Maxwell field equations. With the ansatz of Eq. \eqref{eq:metric}, and in terms of $r_H$, $M$ and $Q$, it reads
\begin{equation}
    \begin{aligned}
        &f_{\mathrm{KN}} = \left(1+\frac{r_H}{r}\right)^2 \frac{\mathcal{F}_1}{\mathcal{F}_2},\\&
        g_{\mathrm{KN}} = \left(1+\frac{r_H}{r}\right)^2,\\&
        h_{\mathrm{KN}} = \frac{\mathcal{F}_1^2}{\mathcal{F}_2},\\&
        W_{\mathrm{KN}} = \frac{r \left(2 M^2-Q^2\right)+2 M \left(r^2+r_H^2\right)}{r_H r^3 \mathcal{F}_2}\sqrt{M^2-Q^2-4 r_H^2}
    \end{aligned}
  \label{eq:kerr-newman}
\end{equation}
where
\begin{equation}
    \begin{aligned}
    &\mathcal{F}_1 = \frac{r^2 \left(2 M^2-Q^2\right)+2 M r \left(r^2+r_H^2\right)+\left(r^2-r_H^2\right)^2}{r^4}-\frac{\left(M^2-Q^2-4 r_H^2\right)}{r^2}\sin^2\theta ,\\&
    \mathcal{F}_2 = \left(\mathcal{F}_1 + \frac{ \left(M^2-Q^2-4 r_H^2\right) }{r^2} \sin^2\theta\right)^2 - \frac{\left(r^2-r_H^2\right)^2 \left(M^2-Q^2-4 r_H^2\right)}{r^6} \sin^2\theta.
    \end{aligned}
\end{equation}
The four-potential functions in Eq. \eqref{eq:ansatzEM} are
\begin{equation}
    A_\varphi = \frac{Q r \left(1 + \frac{M}{r} + \frac{r_H^2}{r^2}\right) \sqrt{M^2-Q^2-4 r_H^2} }{r^2 \left(1 + \frac{M}{r} + \frac{r_H^2}{r^2}\right)^2+\left(M^2-Q^2-4 r_H^2\right)\cos^2\theta},
\end{equation}
and
\begin{equation}
    A_t = \Phi - \frac{Q r \left(1 + \frac{M}{r} + \frac{r_H^2}{r^2}\right)}{r^2 \left(1 + \frac{M}{r} + \frac{r_H^2}{r^2}\right)^2+\left(M^2-Q^2-4 r_H^2\right)\cos^2\theta} + \frac{W_{KN}}{r}\left(1-\mathcal{N}\right) A_\varphi \sin^2 \theta.
\end{equation}
The electrostatic potential $\Phi$ can be chosen such that $A_t|_{r_H} = 0$. This specific parameterization of the functions $A_t$ and $A_\varphi$ for the vector potential is designed to be optimal for a numerical setup similar to the one we are using.

The total angular momentum, $J$, is related to $M$, $Q$ and $r_H$ via
\begin{equation}
  r_H = \frac{1}{2}\sqrt{M^2-\frac{J^2}{M^2}-Q^2} \equiv \frac{M}{2}\sqrt{1-\chi^2-q^2}.
\end{equation}
The Hawking temperature and entropy of a Kerr-Newman black hole are given as
\begin{equation}
    \frac{S}{4\pi M^2} = \frac{1}{2}\left(1 - \frac{q^2}{2} + \sqrt{1-q^2-\chi^2} \right), \qquad 8\pi M T_H = \frac{2 \sqrt{1-q^2-\chi^2}}{1 - \frac{q^2}{2} + \sqrt{1-q^2-\chi^2}}.
\end{equation}
It is worth noting that the Kerr-Newman black hole in the quasi-isotropic coordinate system presented in Eq. \eqref{eq:metric}, can be derived from the standard Boyer-Lindquist coordinate solution through a radial coordinate transformation. The transformation is given by
\begin{equation}
  r_{BL} = r + M + \frac{M^2 - J^2/M^2 - Q^2}{4r} = r \left(1 + \frac{M}{r} + \frac{r_H^2}{r^2}\right).
  \label{eq:rbl}
\end{equation}

\bibliographystyle{ieeetr}
\bibliography{biblio}

\end{document}